\documentclass[aps,prd,reprint,nofootinbib,preprintnumbers,floatfix,superscriptaddress]{revtex4-1}

\pdfoutput=1

\usepackage[bookmarks=false,hyperfootnotes=false]{hyperref}
\usepackage{graphicx}
\usepackage{amsmath}
\usepackage{subfig}
\usepackage{xspace}
\usepackage{slashed}

\usepackage{bbm}

\newcommand{\ecf}[2]{e_{#1}^{(#2)}}

\DeclareRobustCommand{\Sec}[1]{Sec.~\ref{#1}}

\DeclareRobustCommand{\App}[1]{App.~\ref{#1}}

\DeclareRobustCommand{\Fig}[1]{Fig.~\ref{#1}}
\DeclareRobustCommand{\Figs}[2]{Figs.~\ref{#1} and \ref{#2}}
\DeclareRobustCommand{\Eq}[1]{Eq.~(\ref{#1})}

\DeclareRobustCommand{\Ref}[1]{Ref.~\cite{#1}}
\DeclareRobustCommand{\Refs}[1]{Refs.~\cite{#1}}

\newcommand{\pythia}[1]{\textsc{Pythia\xspace #1}}

\newcommand{\fastjet}[1]{\textsc{FastJet\xspace #1}}

\newcommand{\vincia}[1]{\textsc{Vincia\xspace #1}}

\newcommand{\sab}[2]{\langle #1  #2  \rangle}

\begin{document}

\preprint{MIT--CTP 4724}

\title{Non-Global Correlations in Collider Physics}

\author{Andrew J. Larkoski}
\email{larkoski@physics.harvard.edu}
\affiliation{Center for the Fundamental Laws of Nature, Harvard University, Cambridge, MA 02138, USA}
\author{Ian Moult}
\email{ianmoult@mit.edu}
\affiliation{Center for Theoretical Physics, Massachusetts Institute of Technology, Cambridge, MA 02139, USA}

\begin{abstract}
Despite their importance for precision QCD calculations, correlations between in- and out-of-jet regions of phase space have never directly been observed.  These so-called non-global effects are present generically whenever a collider physics measurement is not explicitly dependent on radiation throughout the entire phase space.  In this paper, we introduce a novel procedure based on mutual information, which allows us to isolate these non-global correlations between measurements made in different regions of phase space. We study this procedure both analytically and in Monte Carlo simulations in the context of observables measured on hadronic final states produced in $e^+e^-$ collisions, though it is more widely applicable.  The procedure exploits the sensitivity of soft radiation at large angles to non-global correlations, and we calculate these correlations through next-to-leading logarithmic accuracy.  The bulk of these non-global correlations are found to be described in Monte Carlo simulation. They increase by the inclusion of non-perturbative effects, which we show can be incorporated in our calculation through the use of a model shape function. This procedure illuminates the source of non-global correlations and has connections more broadly to fundamental quantities in quantum field theory.
\end{abstract}

\maketitle

\section{Introduction}

Because of the infrared divergences in a weakly-coupled gauge theory, like QCD at high energy, final states manifest themselves as collections of jets.  To connect the observed final state to final states from fixed-order perturbative calculations, infrared and collinear (IRC) safe jet algorithms \cite{Sterman:1977wj} have been invented to make these jets well-defined.  Traditionally, these jets have been used as a proxy for the short distance degrees of freedom, quarks and gluons, and provide definite objects for matching to fixed-order calculations of cross sections inclusive over the entire final state.  However, with a definite algorithm, individual jets can be studied in their own right, ignoring or integrating over all other radiation in the final state, that is not included in the jet.  This introduces implicit dependence into the jet on the relevant scales of the out-of-jet radiation, the dominant effects of which are referred to as non-global logarithms (NGLs) \cite{Dasgupta:2001sh}.  Non-global effects and logarithms arise generically whenever an observable is only measured on a restricted region of phase space.

For observables measured exclusively on jets, like the jet mass for example, NGLs can introduce large corrections in specific phase space regions, and have proved challenging to understand systematically \cite{Kelley:2011tj,Chien:2012ur,Dasgupta:2012hg,Jouttenus:2013hs}.  There has been significant advances recently in theoretical progress for understanding and calculating NGLs \cite{Dasgupta:2001sh,Dasgupta:2002bw,Dasgupta:2002dc,Banfi:2002hw,Appleby:2002ke,Weigert:2003mm,Rubin:2010fc,Banfi:2010pa,Kelley:2011tj,Hornig:2011iu,Hornig:2011tg,Kelley:2011aa,Kelley:2012kj,Hatta:2013iba,Schwartz:2014wha,Khelifa-Kerfa:2015mma,Caron-Huot:2015bja,Larkoski:2015zka,Hagiwara:2015bia,Becher:2015hka,Neill:2015nya} as well as the development of techniques for eliminating them by removal of appropriate soft radiation in jets \cite{Dasgupta:2013ihk,Dasgupta:2013via,Larkoski:2014wba}.  Despite their importance and relevance, NGLs and non-global correlations in general, have never directly been observed. In particular, their effects on observables such as the jet mass, while potentially large, are difficult to unambiguously disentangle from perturbative and non-perturbative uncertainties when comparing with measurements. In this paper, we introduce a novel procedure which directly measures non-global correlations between in-jet and out-of-jet phase space regions, and which vanishes when such correlations are not present.  We present theoretical calculations of this observable to next-to-leading logarithmic (NLL) accuracy in QCD, and study its properties in Monte Carlo simulation.

While our procedure for measuring non-global correlations can be applied widely, we will restrict our discussion to correlations between hemisphere jets in $e^+e^-$ annihilation to hadrons.  In that case, we first separate events into left and right hemispheres using a cone jet algorithm defined about recoil-free jet axes \cite{Banfi:2004yd,Salambroadening,Larkoski:2014uqa}. This is necessary to eliminate back-reaction of soft particles near the hemisphere boundary on the jet finding.  With identified hemispheres, we then measure IRC safe two-point energy correlation functions $\ecf{2}{\beta}$ \cite{Banfi:2004yd,Jankowiak:2011qa,Gallicchio:2012ez,Larkoski:2013eya}, defined by the angular exponent $\beta$, in each hemisphere.  The energy correlation functions are sensitive to wide-angle, soft radiation and so by measuring the correlation between the measured values in each hemisphere, one defines an observable that is sensitive to non-global correlations.  The problem of measuring non-global correlations is then reduced to defining a measure of the correlation between the left and right hemisphere values of the energy correlation functions.

While there are many measures that can be used to determine the correlation, the one we will use is the mutual information. For the application of non-global correlations, the mutual information measures the number of bits of information that are known about the probability distribution of the right hemisphere energy correlation function, $\ecf{2,R}{\beta}$, given that the distribution of the energy correlation function in the left hemisphere, $\ecf{2,L}{\beta}$, is known (or vice-versa).  The definition of mutual information, $I\left(\ecf{2,L}{\beta},\ecf{2,R}{\beta}\right)$, is
\begin{align}\label{eq:mutinfdef}
I\left(\ecf{2,L}{\beta},\ecf{2,R}{\beta}\right) \\
&
\hspace{-2cm}
= \int d\ecf{2,L}{\beta} \,  d\ecf{2,R}{\beta} \, p\left(\ecf{2,L}{\beta},\ecf{2,R}{\beta}\right)\log_2\frac{p\left(\ecf{2,L}{\beta},\ecf{2,R}{\beta}\right)}{p\left(\ecf{2,L}{\beta}\right)p\left(\ecf{2,R}{\beta}\right)}\,.
\nonumber
\end{align}
Here $p(a)$ is the probability distribution of an observable $a$ and $p(a,b)$ is the joint probability distribution of two observables $a$ and $b$.  Note that 
\begin{equation}
p(a) = \int db\, p(a,b) \,.
\end{equation}
If the left and right hemisphere observables $\ecf{2,L}{\beta}$ and $\ecf{2,R}{\beta}$ are truly uncorrelated, then $$
p\left(\ecf{2,L}{\beta},\ecf{2,R}{\beta}\right)=p\left(\ecf{2,L}{\beta}\right)p\left(\ecf{2,R}{\beta}\right)\,,
$$
and the mutual information is zero.\footnote{\Ref{Larkoski:2014pca} discusses other basic properties of the mutual information and its calculation in the context of correlations of multiple observables measured on the same jet.}  Therefore, the mutual information $I\left(\ecf{2,L}{\beta},\ecf{2,R}{\beta}\right)$ is only non-zero if there are non-trivial correlations between the hemispheres.  Just the absolute value of the mutual information is difficult to interpret because it can range anywhere from 0 to $\infty$.

Mutual information by itself is just one number, and it can potentially be unclear what that number means.  We are able to learn about the origin of the correlation between the left and right hemispheres by measuring the mutual information over a range of angular exponents $\beta$.  As we will discuss in \Sec{sec:mutinfcalc}, the energy correlation functions are defined such that by increasing the angular exponent $\beta$, one becomes increasingly sensitive to soft, wide angle emissions.  These soft wide angle emissions are the most sensitive to non-global correlations between the hemispheres.  Therefore, as the angular exponent $\beta$ increases, $I\left(\ecf{2,L}{\beta},\ecf{2,R}{\beta}\right)$ should also increase due to the increasing importance of soft emissions sensitive to non-global correlations.  As the $\beta$ dependence of $I\left(\ecf{2,L}{\beta},\ecf{2,R}{\beta}\right)$ is most important for understanding non-global effects, we will denote $I\left(\ecf{2,L}{\beta},\ecf{2,R}{\beta}\right)\equiv I_\beta$.  Monotonically increasing $I_\beta$ with $\beta$ is an unambiguous signal of non-global physics.  There may be baseline correlations due to other effects, but these would not be manifest as a rise in $I_\beta$ with $\beta$.

An important prediction of this framework is that when appropriate methods are used to remove soft wide-angle radiation, so-called jet grooming algorithms, the correlations between the hemispheres should vanish.  It has been shown that the modified mass drop \cite{Dasgupta:2013ihk,Dasgupta:2013via} and the soft drop \cite{Larkoski:2014wba} jet groomers eliminate NGLs, at least through NLL accuracy.  Therefore, in our analytical calculations, when these jet grooming techniques are applied, $I_\beta$ is close to zero and does not increase with $\beta$.  This is also borne out in Monte Carlo simulation of $e^+e^-$ collision events.

We choose to use mutual information to measure correlations as it has been used in some other studies of correlations of QCD jet observables \cite{Carruthers:1989gu,Narsky:2014fya,Larkoski:2014pca}.  However, any procedure that tests the difference between the joint probability distribution of two observables and the product of the individual probability distributions would be sufficient.  Other examples, though not exhaustive, that quantify correlations used in statistical analyses include Hellinger distance, Jensen-Shannon divergence, Kolmogorov-Smirnov test, and R\'enyi divergence.  The Hellinger distance is actually a metric and ranges strictly between 0 and 1. More ambitiously, although we have focused on methods which allow for the reduction of the correlations to a single number, it would also be interesting to directly study the double differential distribution of $p\left(\ecf{2,L}{\beta},\ecf{2,R}{\beta}\right)-p\left(\ecf{2,L}{\beta}\right)p\left(\ecf{2,R}{\beta}\right)$. This is, however, considerably more complicated to directly interpret, and we therefore restrict ourselves to the study of the mutual information, which summarizes the correlations in a single number.

It is tempting to draw an analogy between these non-global correlations and quantum entanglement of the two hemispheres, but mutual information is measured on the asymptotic states, long after decoherence.  It is therefore challenging to make a direct connection to the underlying quantum correlations.  More generally, in an experiment that only measures energy deposits it is not possible to construct non-commuting operators at the same point in the detector, so as to test Bell's inequalities \cite{Bell:1964kc}.  It may be possible to do so, however, with idealized detectors in toy models, as was recently proposed in the context of inflation \cite{Maldacena:2015bha}.  We leave a study of possible connections between NGLs and entanglement to future work.

The outline of this paper is as follows.  In \Sec{sec:ngp}, we review the source of dominant non-global correlations and logarithms from fixed-order matrix elements in $e^+e^-$ collisions.  In \Sec{sec:mutinfcalc}, we define the two-point energy correlation functions appropriate for $e^+e^-$ collisions and calculate the mutual information $I_\beta$ to NLL accuracy, capturing non-global logarithms with the dressed gluon expansion of \Ref{Larkoski:2015zka}.  We are also able to include non-perturbative effects by convolving the perturbative distributions with a shape function, and we demonstrate that non-perturbative effects in general increase the correlation between the hemispheres.  In \Sec{sec:mc}, we calculate the mutual information in Monte Carlo simulation and find good agreement between our analytical calculation and the simulations.  We conclude in \Sec{sec:conc} and discuss other scenarios where non-global effects can be directly measured in a similar manner.  Calculational details both of the analytics and algorithms for determining the mutual information on finite data sets are presented in appendices.

\section{Non-Global Physics}\label{sec:ngp}

In this section, we briefly review non-global physics in the context of $e^+e^-\to$ dijets events.  Our goal is to calculate the distribution of observables measured on the left and right hemispheres (appropriately defined) of each event and we choose these observables such that a non-zero value regulates the soft and collinear singularities.  While the two-point energy correlation functions are one example of such observables, other examples include hemisphere masses or hemisphere thrust.  By demanding that both hemisphere masses are non-zero, for example, means that the lowest order contribution to the double differential cross section or joint probability distribution is at ${\cal O}(\alpha_s^2)$.  Sufficient for our analysis here, we can consider purely gluonic radiation from the initial $q\bar q$ dipole.  Then, we must calculate the Feynman diagrams that contribute to this final state.

If we assume that QCD is an abelian gauge theory, then gluons are emitted exclusively off of the $q\bar q$ dipole.  For two positive helicity abelian gluons, the matrix element can be written in spinor helicity notation \cite{Mangano:1990by,Dixon:1996wi} as
\begin{equation}\label{eq:abelme}
\frac{{\cal A}(e_a^-, \bar e_b^+\to q_1^+,g_2^+,g_3^+,\bar q_4^-) }{{\cal A}(e_a^-, \bar e_b^+\to q_1^+,\bar q_4^-)}=\frac{\sab{1}{4}}{\sab{1}{2}\sab{2}{4}}\frac{\sab{1}{4}}{\sab{1}{3}\sab{3}{4}}\,,
\end{equation}
where the quarks are particles $1$ and $4$, the gluons are particles $2$ and $3$, and we have stripped away couplings.  The Born-level matrix element is
\begin{equation}
{\cal A}(e_a^-, \bar e_b^+\to q_1^+,\bar q_4^-) = \frac{\sab{4}{a}^2}{\sab{1}{4}\sab{a}{b}}\,,
\end{equation}
again, stripped of couplings, with the initial electron and positron particles $a$ and $b$.  The spinor products are defined with $|\sab{i}{j}|^2=s_{ij}$.  In \Eq{eq:abelme}, we have dropped the momentum conserving $\delta$-function, also.  Momentum conservation does correlate the two hemispheres, but in the limit that gluons $2$ and $3$ become soft, this correlation vanishes.  Note also that even away from the soft limit the matrix element factorizes into two contributions, one for each gluon.  In the soft limit and demanding that the gluons are in different hemispheres, the gluons are completely uncorrelated.  Therefore, in an abelian gauge theory like QED, there are no NGLs to this order in perturbation theory.

In a non-abelian gauge theory, however, gluons can radiate more gluons.  For the same selection of helicities, the color-ordered matrix element for the emission of two gluons is
\begin{align}
\frac{{\cal A}(e_a^-, \bar e_b^+\to q_1^+,g_2^+,g_3^+,\bar q_4^-) }{{\cal A}(e_a^-, \bar e_b^+\to q_1^+,\bar q_4^-)}= \frac{\sab{1}{4}}{\sab{1}{2}\sab{2}{3}\sab{2}{4}} \,,
\end{align}
where, again, we strip couplings and momentum-conserving $\delta$-functions.  Taking the soft limit of gluons 2 and 3 removes correlations through momentum conservation, but even in this limit, the matrix element does not factorize.  The soft gluons in each hemisphere will know about the other through the matrix element, unlike in an abelian theory.  Therefore, in the soft limit, these correlations will manifest themselves as large logarithms in the cross section of the hemisphere observables.  Resummation of these large logarithms is necessary for convergence of the perturbation theory in the singular region of phase space.  In the next section, we present the details of the method for measuring these correlations directly.

\section{Analytical Calculation of Mutual Information}\label{sec:mutinfcalc}

Isolating non-global correlations between the hemispheres is subtle and robust definitions of the hemispheres and observables measured on them are required.  In this section, we define how we identify hemispheres, measure the energy correlation functions, and present a calculation to NLL accuracy.

\subsection{Observable Definitions}

Because wide angle soft radiation is most sensitive to non-global correlations, we need a procedure for identifying the hemispheres that is insensitive to back-reaction by the soft particles on the boundary.  Effectively, we must define the hemispheres with an algorithm that does not introduce clustering logarithms \cite{Banfi:2005gj,Delenda:2006nf}.  Our procedure for doing this is the following.  We first cluster the event with the exclusive $k_T$ algorithm \cite{Catani:1991hj} to two jets using Winner-Take-All (WTA) recombination \cite{Salambroadening,Larkoski:2014uqa,Larkoski:2014bia}.  WTA recombination ensures that the jet axes lie along the direction of the hardest radiation in the event, and are not displaced by recoil effects from soft, wide angle radiation.  We then identify the axis defined in this way of the highest energy jet in the event and cluster all radiation within an angle of $\pi/2$ of this axis into one hemisphere, and all of the remaining radiation into the other hemisphere.  This effectively defines an exclusive cone jet algorithm.  We then randomly choose one of the hemispheres to be the left hemisphere, and the other to be the right hemisphere.

With these hemisphere definitions, we then measure energy correlation functions on each hemisphere.  The two-point energy correlation functions $\ecf{2}{\beta}$ are defined for $e^+e^-$ collision events as
\begin{equation}
\ecf{2}{\beta} = \frac{1}{E_J^2}\sum_{i< j \in J} E_i E_j\left(
\frac{2p_i\cdot p_j}{E_i E_j}
\right)^{\beta / 2}\,,
\end{equation}
where $E_J$ is the total energy of the jet (or otherwise identified region), the sum runs over all distinct pairs of particles in the jet, and $\beta$ is an angular exponent required to be larger than 0 by IRC safety.  Note that for massless particles,
\begin{equation}
\frac{2p_i\cdot p_j}{E_i E_j} = 2(1-\cos\theta_{ij}) \,,
\end{equation}
and so larger values of $\beta$ give greater weight to wide-angle emissions.  The two-point energy correlation function is non-zero for a jet with two particles.  By separately measuring $\ecf{2}{\beta}$ in each hemisphere and demanding that both are non-zero requires that each hemisphere has at least two particles.  We will denote the energy correlation functions measured on the left (right) hemisphere as $\ecf{2,L}{\beta}$ ($\ecf{2,R}{\beta}$).

We emphasize that no other cuts are performed: the left and right hemisphere energy correlation functions are measured on every $e^+e^-\to$ hadrons collision event.

\subsection{Perturbative Calculation}

With robust jet definitions and observables measured on the jet hemispheres, we now calculate the joint probability distribution or double differential cross section of the measured values of $\ecf{2,L}{\beta}$ and $\ecf{2,R}{\beta}$.  Our calculation will be accurate to NLL order, and here, we will not include fixed-order corrections.  These are expected to be a small effect on our results because the bulk of the distribution is in the region where logarithms of the observables are large, and therefore well-described by a resummed calculation.

The first relevant fixed-order corrections arise at ${\cal O}(\alpha_s^2)$, when each hemisphere has an extra emission.  The ${\cal O}(\alpha_s)$ corrections are not representative of power corrections, because one hemisphere necessarily has zero mass.  That means that at ${\cal O}(\alpha_s)$, the hemispheres are completely uncorrelated (except for total momentum conservation).  Matching fixed-order at ${\cal O}(\alpha_s^2)$ to our resummed expressions will be addressed in future work.

For the resummation of energy correlation functions measured in each hemisphere, there are two components that contribute to NLL accuracy.  First, there are global logarithms, which arise from soft or collinear emissions in each hemisphere that are uncorrelated with emissions in the other hemisphere.  In \App{app:resum}, we present the calculation of the global logarithms of the two-point energy correlation functions measured on a hemisphere to NLL accuracy using soft-collinear effective theory (SCET) \cite{Bauer:2000yr,Bauer:2001ct,Bauer:2001yt,Bauer:2002nz}, though other methods can be used and produce identical results \cite{Banfi:2004yd}.  The global cumulative probability distribution or double cumulative cross section $\Sigma^G$ for the hemisphere energy correlation functions can be expressed as
\begin{equation}
\Sigma^\text{G}\left(\ecf{2,L}{\beta},\ecf{2,R}{\beta}\right)\equiv \Sigma\left(\ecf{2,L}{\beta}\right)\Sigma\left(\ecf{2,R}{\beta}\right)\,,
\end{equation} 
where $\Sigma\left(\ecf{2,L}{\beta}\right)$ is the global NLL resummed cumulative distribution of the left-hemisphere energy correlation function.  Because the left and right hemispheres are otherwise identical, $\Sigma\left(\ecf{2,L}{\beta}\right)$ is identical to $\Sigma\left(\ecf{2,R}{\beta}\right)$.  If this was the only contribution to the distribution, then the mutual information of the two hemispheres would be zero, as there is no correlation between between the hemispheres.

However, in addition to global logarithms, at NLL accuracy, there are also NGLs that must be included that introduce correlations between the hemispheres.  To this accuracy, NGLs can be included by Monte Carlo simulation \cite{Dasgupta:2001sh} or by solving the Banfi-Marchesini-Smye (BMS) equation \cite{Banfi:2002hw}.  Here, we will use the dressed gluon expansion \cite{Larkoski:2015zka}, including NGLs with the one-dressed gluon approximation.  The one-dressed gluon does not include complete NGLs at NLL accuracy, but was shown to agree with the solution of the BMS equation at the percent level over the range
$$
\frac{\alpha_s C_A}{\pi} \left|\log\frac{\ecf{2,R}{\beta}}{\ecf{2,L}{\beta}}\right| \lesssim 1.5\,.
$$
The one-dressed gluon resummation of non-global logarithms can be expressed as
\begin{align}
S^{(1,\text{NGL})}_{n\bar n}(\mu_L,\mu_R) &= 1-\frac{2 C_F}{\beta_0}\left[
\gamma_E \left| \log\frac{\alpha_s(\mu_R)}{\alpha_s(\mu_L)}\right|\right.\\
&
\hspace{-1cm}
\left.+\frac{\beta_0}{2 C_A}\log \Gamma\left(
1+\frac{2 C_A}{\beta_0}\left| \log\frac{\alpha_s(\mu_R)}{\alpha_s(\mu_L)}\right|
\right)
\right]\ .\nonumber
\end{align}
Here, $\Gamma(x)$ is Euler's Gamma function, $\gamma_E\simeq 0.577$ is the Euler-Mascheroni constant, and $\beta_0$ is the leading coefficient of the $\beta$-function,
\begin{equation}
\beta_0 = \frac{11}{3}C_A -\frac{2}{3}n_f\,,
\end{equation}
where $n_f$ is the number of active quark flavors.  The scales $\mu_L$ and $\mu_R$ are set by the measured values of the left and right hemisphere energy correlation functions.  Up to factors of order 1, the left hemisphere scale is, for example,
\begin{equation}
\mu_L \simeq \ecf{2,L}{\beta} E_L\,,
\end{equation}
where $E_L$ is the energy of the left hemisphere.  This scale can be varied to provide a measure of perturbative uncertainties.

Then, to NLL accuracy with the one-dressed gluon to capture NGLs, the double cumulative cross section of the hemisphere energy correlation functions can be expressed as
\begin{equation}
\Sigma\left(\ecf{2,L}{\beta},\ecf{2,R}{\beta}\right) =\Sigma\left(\ecf{2,L}{\beta}\right)\Sigma\left(\ecf{2,R}{\beta}\right)S^{(1,\text{NGL})}_{n\bar n}(\mu_L,\mu_R)\,.
\end{equation}
The double differential cross section or joint probability distribution is found by the double derivative:
\begin{equation}\label{eq:pertcalc}
p\left(\ecf{2,L}{\beta},\ecf{2,R}{\beta}\right)=\frac{1}{\sigma}\frac{\partial^2}{\partial \ecf{2,L}{\beta}\, \partial \ecf{2,R}{\beta}} \Sigma\left(\ecf{2,L}{\beta},\ecf{2,R}{\beta}\right) \,.
\end{equation}
With this result, we can then calculate the mutual information $I_\beta$ as defined in \Eq{eq:mutinfdef}.\footnote{While other non-global observables have appeared in the literature e.g.~that of \Refs{Bertolini:2013iqa,Bertolini:2015pka}, a key advantage of the use of mutual information between IRC safe observables is that we have a precise definition of the non-global correlations that are probed in terms of an all orders factorization theorem. The above discussion defines precisely what this constitutes for the case of hadronic jets in $e^+e^-$.}

\begin{figure}[t]
\includegraphics[width=7.5cm]{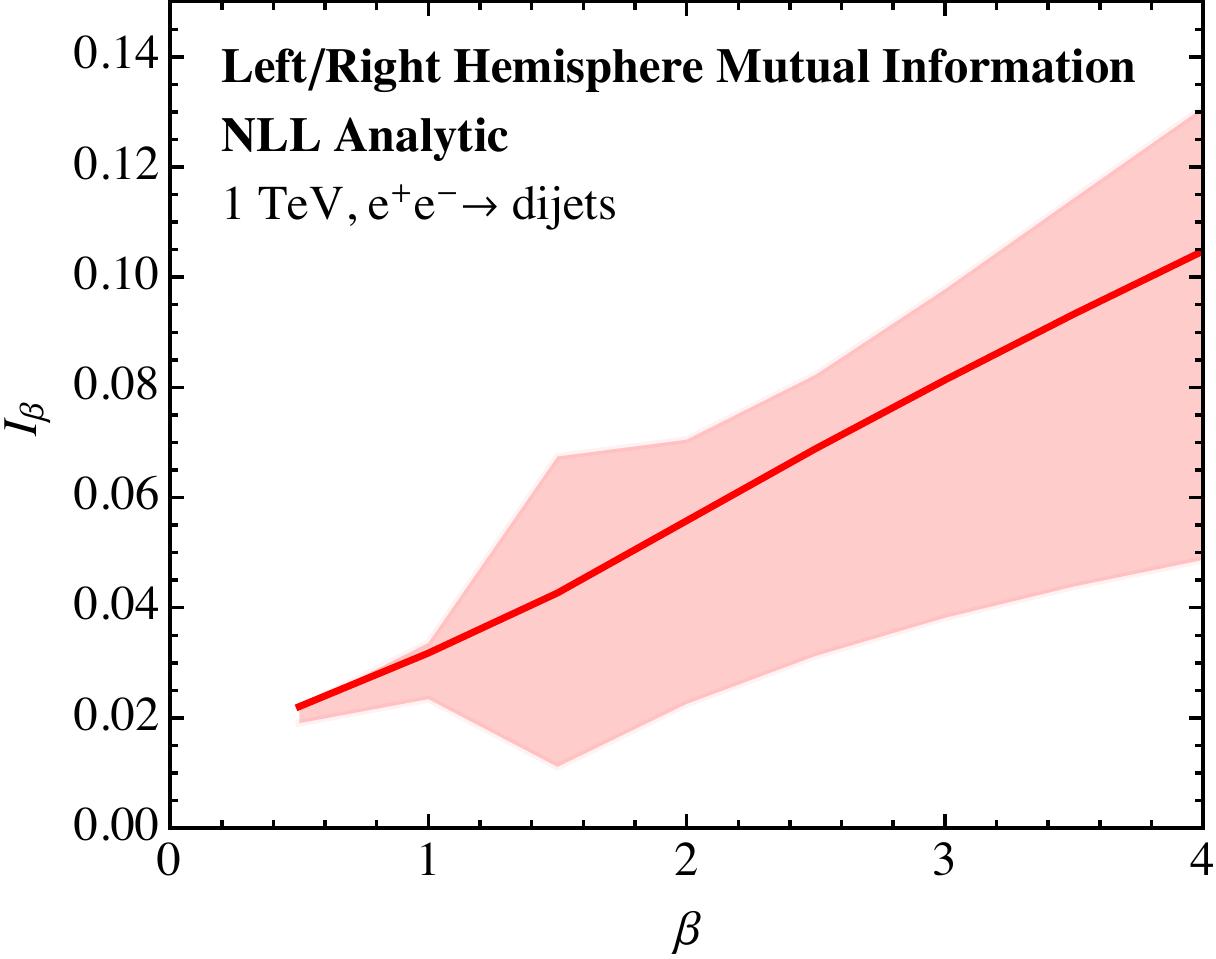}    
\caption{The mutual information $I_\beta$ as calculated to NLL accuracy with the one-dressed gluon to capture NGLs.  The lighter band reflects conservative theoretical uncertainties.
}
\label{fig:pert_mi}
\end{figure}

In \Fig{fig:pert_mi}, we plot the mutual information $I_\beta$ as calculated from \Eq{eq:pertcalc} as a function of the angular exponent $\beta$ for $e^+e^-$ collisions at 1 TeV center of mass energy.  As expected, the mutual information is non-zero and increases with $\beta$, reflecting the increasing importance of soft, wide angle emissions to the energy correlation functions.  The lighter band is representative of theoretical uncertainties, determined by varying the natural scales appearing in the double differential cross section by factors of 2 and taking the envelope.  While the uncertainties are large, the increase of $I_\beta$ with $\beta$ is robust.

\begin{figure}[t]
\includegraphics[width=7.5cm]{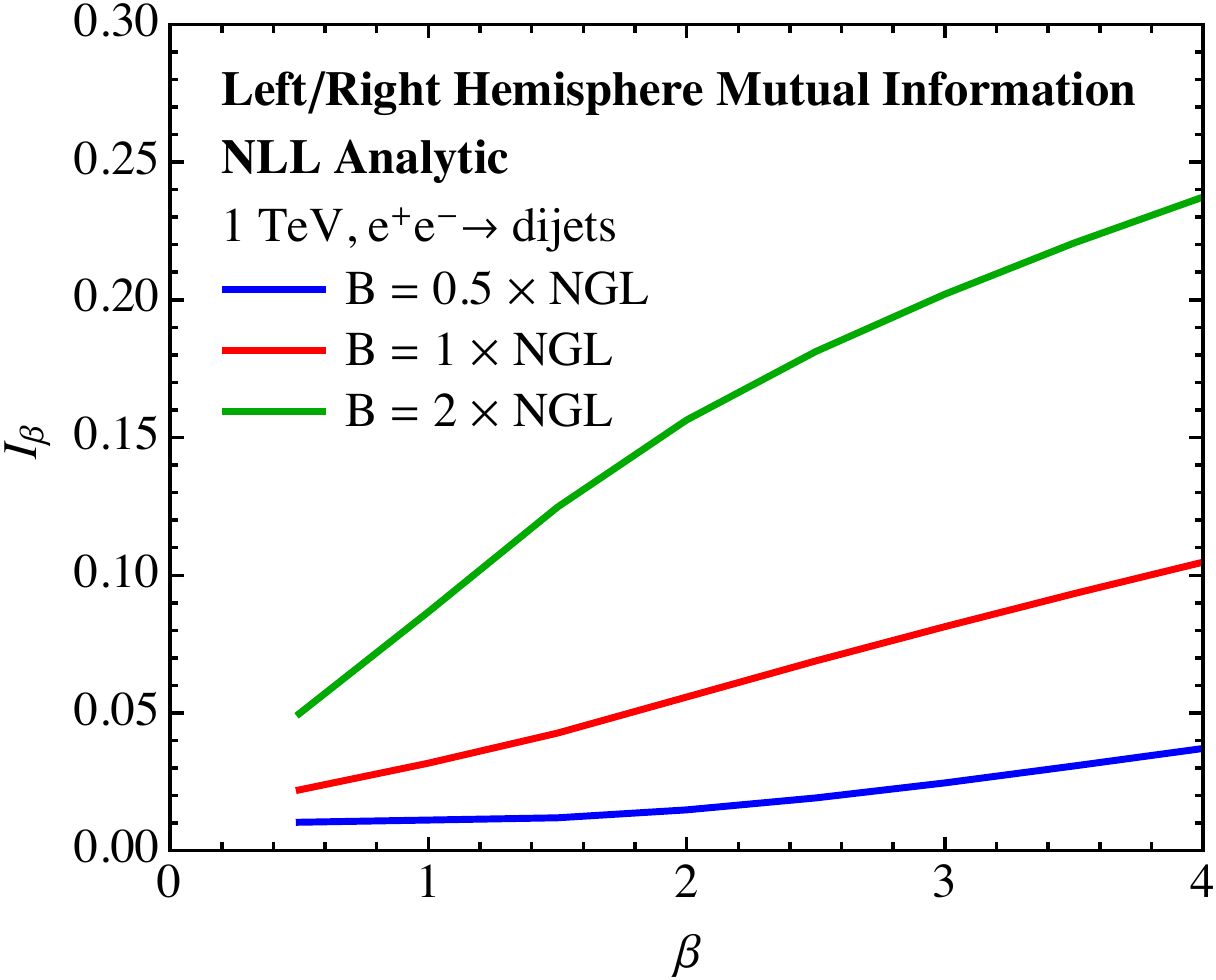}    
\caption{Plot of the dependence of the mutual information on the size of the non-global contributions.  The parameter $B$ controlling the size of the NGLs is varied from $0.5$ to $2$.
}
\label{fig:pert_mi_B}
\end{figure}

It is interesting to study the sensitivity of the mutual information $I_\beta$ to the size of the NGLs.  We can demonstrate this sensitivity by modifying the one-dressed gluon by a coefficient $B$ to be
\begin{align}
\hspace{-0.2cm}S^{(1,\text{NGL})}_{n\bar n}(\mu_L,\mu_R;B) &= 1-B\frac{2 C_F}{\beta_0}\left[
\gamma_E \left| \log\frac{\alpha_s(\mu_R)}{\alpha_s(\mu_L)}\right|\right.\\
&
\hspace{-0.75cm}
\left.+\frac{\beta_0}{2 C_A}\log \Gamma\left(
1+\frac{2 C_A}{\beta_0}\left| \log\frac{\alpha_s(\mu_R)}{\alpha_s(\mu_L)}\right|
\right)
\right]\ .\nonumber
\end{align}
By varying $B$ we can observe the corresponding response of the mutual information.  In \Fig{fig:pert_mi_B}, we plot the mutual information $I_\beta$ for $B = 0.5,\, 1,\, 2$, without including theoretical uncertainties.  $I_\beta$ exhibits roughly linear dependence on $B$, demonstrating that this observable is very sensitive to both the value of $\alpha_s$ and the size of non-global effects.

\subsection{Including Non-Perturbative Effects}

One can additionally include the effects of non-perturbative physics due to hadronization by convolution with a non-perturbative shape function \cite{Korchemsky:1999kt,Korchemsky:2000kp} because the energy correlation functions are additive observables.  Korchemsky and Tafat \cite{Korchemsky:2000kp} introduced a shape function differential in both hemisphere scales $\epsilon_L$ and $\epsilon_R$, $F(\epsilon_L,\epsilon_R)$.  The non-perturbative distribution can then be expressed as
\begin{align}
\sigma(\ecf{2,L}{\beta},\ecf{2,R}{\beta}) &\\
&
\hspace{-1.5cm}
=\int d\epsilon_L\, d\epsilon_R\, F(\epsilon_L,\epsilon_R)\, \sigma_p\left( \ecf{2,L}{\beta} - \frac{\epsilon_L}{E_L},\ecf{2,R}{\beta}-\frac{\epsilon_R}{E_R} \right)\,, \nonumber
\end{align}
where $\sigma_p$ denotes the perturbative distribution.  The shape function is normalized:
\begin{equation}
1=\int d\epsilon_L\, d\epsilon_R\, F(\epsilon_L,\epsilon_R) \,,
\end{equation}
and has support over a region of size set by the non-perturbative scale of QCD, $\Lambda_\text{QCD}$.  The parametrization of the shape function introduced by Korchemsky and Tafat is
\begin{equation}\label{eq:tafat_korchemsky}
F(\epsilon_L,\epsilon_R) = N \left(\frac{\epsilon_L\epsilon_R}{\Lambda^2}\right)^{a-1}e^{-\frac{\epsilon_L^2+\epsilon_R^2+2b\epsilon_L\epsilon_R}{\Lambda^2}}\,.
\end{equation}
$N$ is set by the normalization of the shape function, and $a$, $b$, and $\Lambda$ are parameters of the shape function.  By fitting data for heavy jet mass, they suggested the values $a=2$, $b=-0.4$, and $\Lambda = 0.55$ GeV. It is important to emphasize that this is a parameterization, and other possibilities are possible. Most generally, a complete basis of shape functions should be considered \cite{Ligeti:2008ac}. 

The parameter $b$ controls the amount of correlation between the hemispheres introduced by non-perturbative effects.  A value of $b=0$ eliminates non-perturbative correlations, while $b>-1$ is required for the shape function to be normalizable.  In general, we expect the amount of non-perturbative correlation to depend on the angular exponent $\beta$ of the energy correlation functions.  We are able to parametrize this dependence by considering simple limiting cases.  As $\beta\to \infty$, the non-perturbative correlations should become unbounded because the mutual information is controlled by emissions right at the hemisphere boundary.  Therefore, as $\beta \to \infty$, we expect that $b \to -1$.  If $\beta \leq1$, then collinear emissions live at an equal or lower virtuality than soft emissions.  Thus, for $\beta \leq 1$, non-perturbative corrections will be dominated by high energy collinear splittings with relative transverse momentum near the scale $\Lambda_\text{QCD}$.  The non-perturbative correlations should therefore vanish for $\beta \lesssim 1$, because different hard collinear sectors are uncorrelated, up to corrections at higher powers in the observables.  Therefore, as $\beta \to1$, $b\to 0$.  A simple parametrization of $b$ that accounts for these expected limits is
\begin{equation}
b = \frac{1-\beta}{\beta} \,.
\end{equation}
We do not, however, have a proof of this relation. Measurement of $b$ as a function of $\beta$, perhaps using our mutual information observable, would shed light onto the non-perturbative correlations between observables.
For the hemisphere mass, $\beta=2$ and $b=-0.5$, close to the value suggested by Korchemsky and Tafat.  We will therefore use this value of $b$ in our analysis.  Since our perturbative uncertainties are large, we will not study in detail the variation of non-perturbative parameters. Instead, our focus is simply on showing how non-perturbative effects can be incorporated, and understanding their impact on the mutual information.

It is interesting to note that while to leading logarithmic accuracy the mutual information, $I_\beta$, vanishes in perturbation theory, this is no longer true once non-perturbative effects are included with a non-zero value of the shape function parameter $b$. A non-zero $b$ induces correlations between the energy correlation functions as measured on the left and right hemispheres even if they are not present perturbatively. A measurement of the mutual information could therefore also prove useful in constraining non-perturbative correlations in event shape observables, and improving their modeling in Monte Carlo programs.

The non-perturbative scale $\Lambda$ will also have dependence on $\beta$.  From the universality of the leading non-perturbative corrections \cite{Dokshitzer:1995zt,Lee:2006nr}, it has been demonstrated that for observables like the energy correlation functions that are additive and have an angular exponent parameter, the non-perturbative scale $\Lambda$ is
\begin{equation}
\Lambda = \frac{\Omega}{\beta - 1} \,,
\end{equation}
where $\Omega$ is a fixed energy scale set by a non-perturbative matrix element.  Following Korchemsky and Tafat, we set $\Omega = 0.55$ GeV.  Note that this is only sensible for $\beta > 1$, where soft emissions dominate the non-perturbative corrections.

While non-perturbative correlations can be large, they do not necessarily result in a large change of the mutual information.  Non-perturbative corrections are only dominant near the singular regions of phase space.  For large values of the energy correlation functions the effect of the shape function reduces to a shift of the perturbative distribution by an amount proportional to the non-perturbative scale.  However, a shift of the distribution does not affect the mutual information (which can be seen by a simple change of variables in its definition).  So, non-perturbative effects contribute to the mutual information an amount suppressed by the volume of phase space in which they dominate.

\begin{figure}[t]
\includegraphics[width=7.5cm]{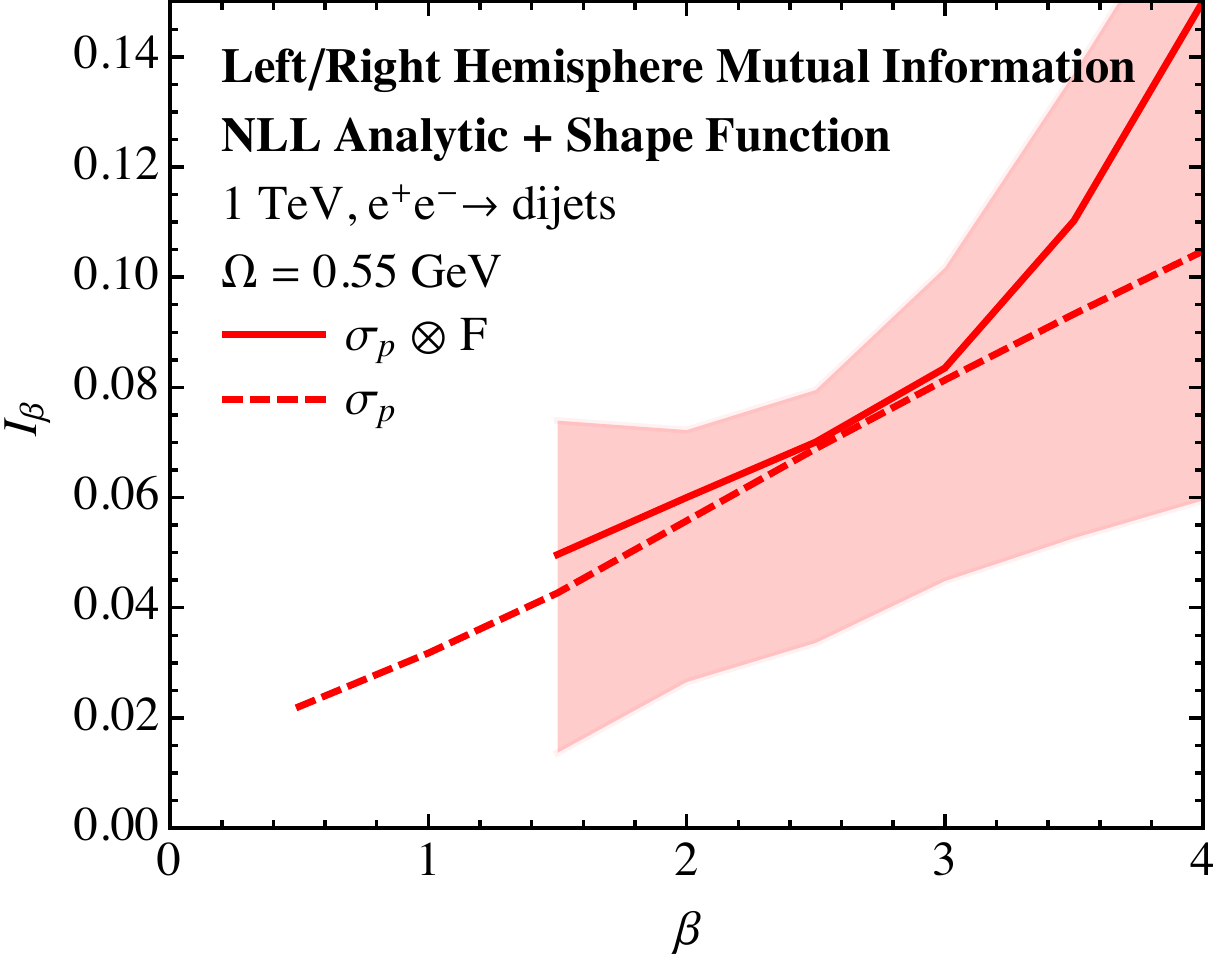}    
\caption{The mutual information $I_\beta$ as calculated to NLL accuracy with the one-dressed gluon to capture NGLs including a non-perturbative shape function (solid).  The lighter band reflects conservative theoretical uncertainties and the purely perturbative result is shown for comparison (dashed).
}
\label{fig:pert_mi_np}
\end{figure}

In \Fig{fig:pert_mi_np}, we plot the mutual information $I_\beta$ including non-perturbative corrections, as a function of $\beta$.  For this plot, we set the parameters in the shape function to be
\begin{align*}
a&= 2\,,\\
b&=\frac{1-\beta}{\beta} \,, \\
\Lambda &= \frac{0.55\ \text{GeV}}{\beta -1} \,.
\end{align*}
Additionally, we include an estimate of perturbative uncertainties by varying scales in the perturbative distribution.  For comparison, we also include the perturbative mutual information.  We can only compute the non-perturbative mutual information for $\beta \gtrsim 1$, for the reasons described above.  Especially as $\beta$ increases, we see that non-perturbative correlations further increase $I_\beta$, as expected.  While we only plot for one value of the non-perturbative parameter $\Lambda$, we found only very weak dependence on this value.

\section{Monte Carlo Study}\label{sec:mc}

\begin{figure*}[t]
\centering
\subfloat[]{\label{fig:nsub_sb1}
\includegraphics[width=7.5cm]{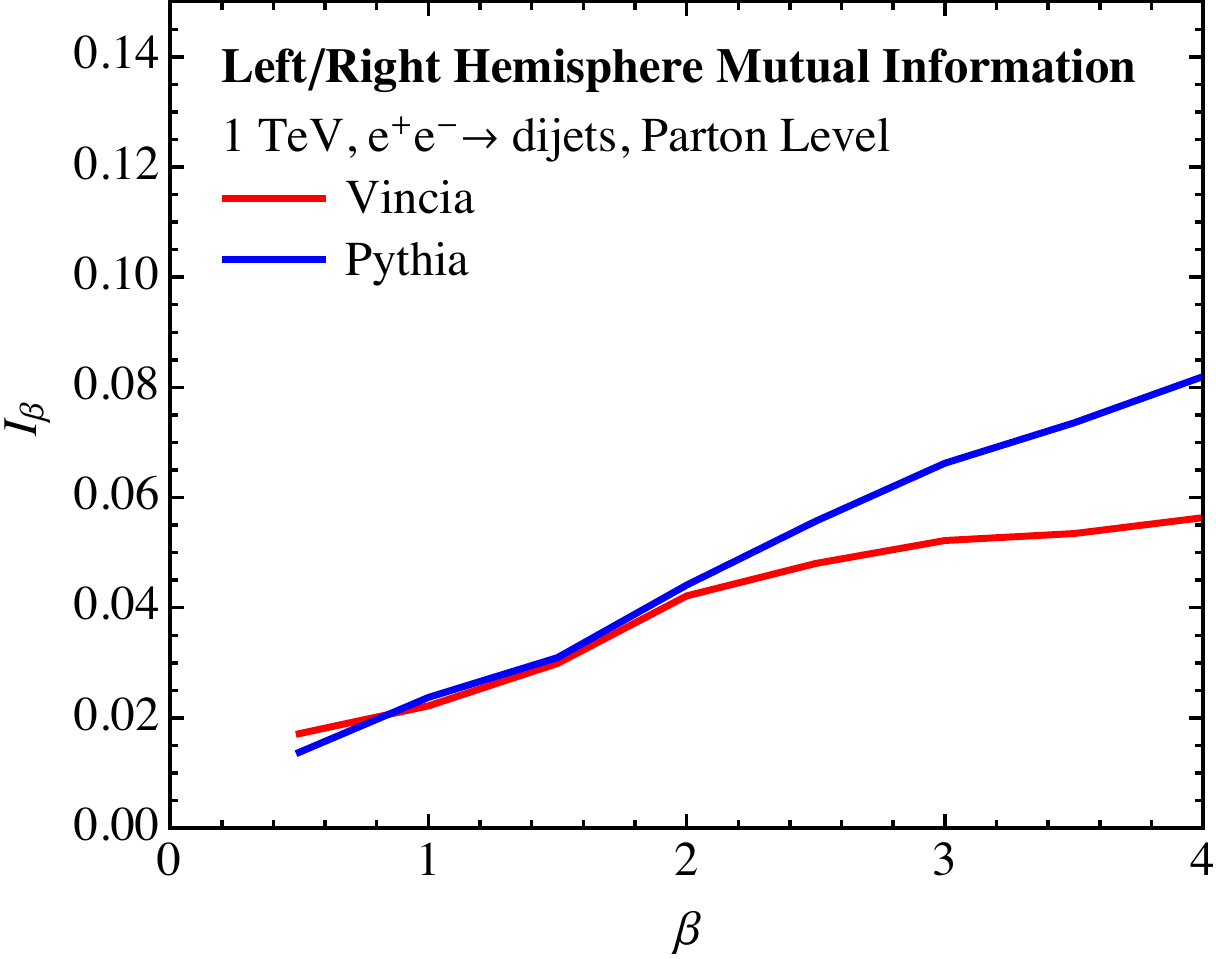}    
} \qquad \qquad
\subfloat[]{\label{fig:nsub_sb2}
\includegraphics[width=7.4cm]{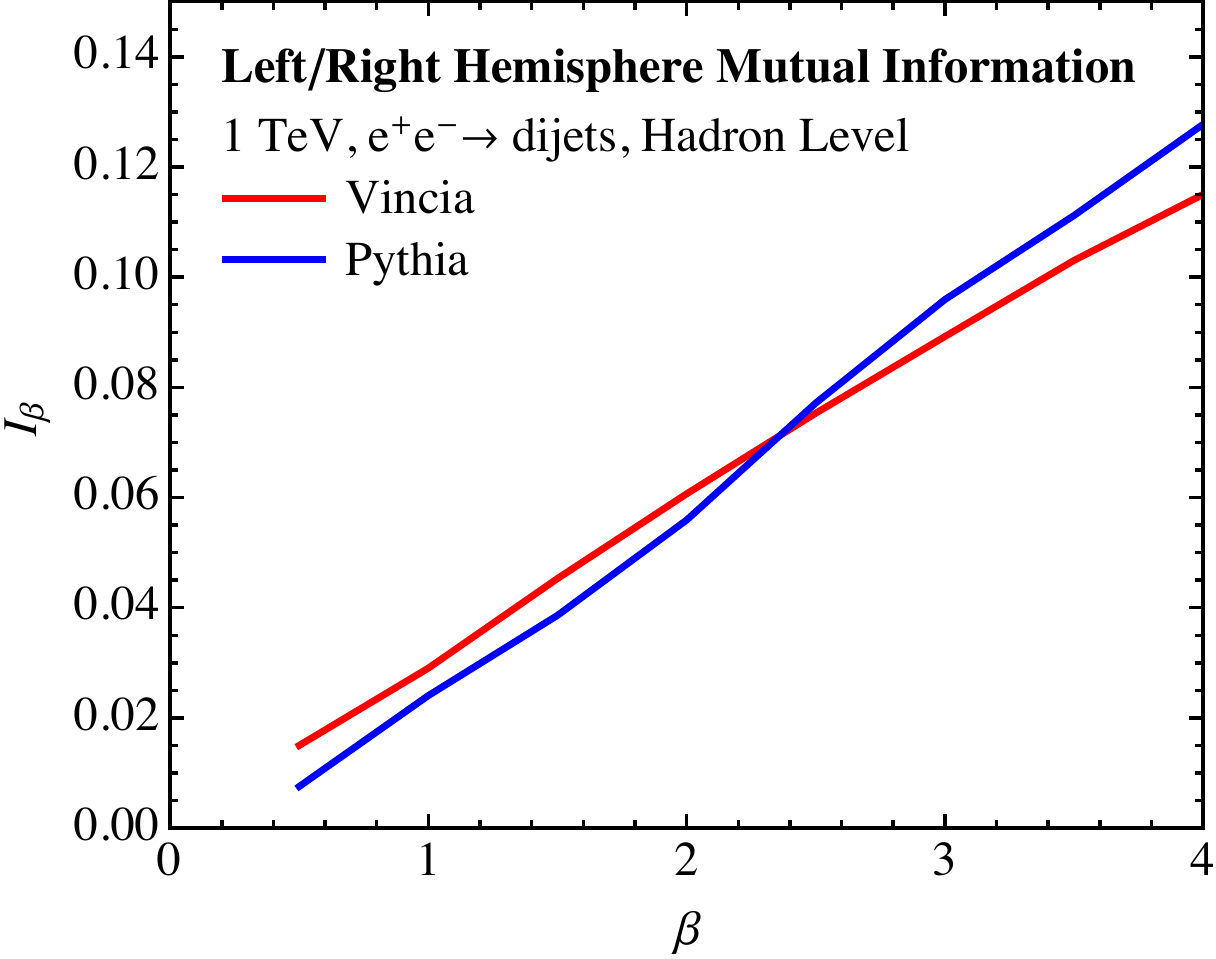} 
}\\
\caption{ The mutual information $I_\beta$ as calculated in \pythia{} and \vincia{} Monte Carlo at (a) parton-level and (b) hadron-level.
}
\label{fig:mi_MC}
\end{figure*}

In this section, we compare our analytical calculations of non-global correlations to Monte Carlo simulation of $e^+e^-$ collision events.  We generate $e^+e^-$ collision events at 1 TeV center of mass energy with both \pythia{8.210}  \cite{Sjostrand:2007gs,Sjostrand:2014zea} and \vincia{1.2.02} \cite{Giele:2007di,Giele:2011cb,GehrmannDeRidder:2011dm,Ritzmann:2012ca,Hartgring:2013jma,Larkoski:2013yi} Monte Carlos.  For comparison to our analytic calculations, we consider parton level and hadron level events.  We then cluster the event into two jets with the exclusive $e^+e^-$ $k_T$ algorithm \cite{Catani:1991hj} with WTA recombination \cite{Larkoski:2014uqa,Larkoski:2014bia,Bertolini:2013iqa,Salambroadening} as implemented in \fastjet{3.1.3} \cite{Cacciari:2011ma}.  About the axis of the highest energy jet in the event, we cluster particles into one hemisphere if they lie within an angle of $\pi/2$ of this axis, and cluster them into the other hemisphere if they are outside this region.  The hemispheres are then randomly assigned to be left or right and the energy correlation functions are measured on them.

The calculation of the mutual information from these event samples is quite subtle, and a detailed discussion of calculating mutual information from finite statistics is described in \Ref{Larkoski:2014pca}.\footnote{The effects described there have actually been known for a very long time in other applications of mutual information; see, e.g., \Refs{treves1995upward,panzeri1996analytical}.}  While we do not present a detailed discussion of the calculation of mutual information on finite data, we will present the results.  Mutual information can be equivalently defined through the Shannon entropies of the various distributions.  For observables $a$ and $b$, the mutual information can be written as
\begin{equation}
I(a,b) = H(a)+H(b)-H(a,b)\,.
\end{equation}
For binned data, the entropy $H(a)$ is
\begin{equation}
H(a) = -\sum_{i\in \text{bins}}\frac{n_i}{N^{a}_\text{ev}}\log_2\frac{n_i}{N_\text{ev}}\,,
\end{equation}
where $N^a_\text{ev}$ is the total number of events in the $a$ sample and $n^a_i$ is the number of events in bin $i$ for observable $a$.  For a finite data set, the number of events per bin will fluctuate, and the leading effect of these fluctuations on the mutual information can be calculated.  The mutual information is then
\begin{align}\label{eq:mutinf_fluc}
I(a,b) = I_\infty(a,b)- \frac{1}{2\log 2} \left(  \frac{n^a_\text{bins}}{N^{a}_\text{ev}} + \frac{n^b_\text{bins}}{N^{b}_\text{ev}} - \frac{n^{ab}_\text{bins}}{N^{ab}_\text{ev}}  \right) +\dotsc
\end{align}
Here, $I_\infty(a,b)$ is the mutual information with infinite statistics, $n^a_\text{bins}$ is the number of bins in the $a$ sample, and $N^{a}_\text{ev}$ is the number of events in the $a$ sample.  To remove, or at least minimize, effects of finite binning, we want to set the term in parentheses in \Eq{eq:mutinf_fluc} equal to zero with judicious choices of bin and sample sizes.  The choice used in this paper is to set $n^a_\text{bins}=n^b_\text{bins}\equiv n_\text{bins}$ and $n^{ab}_\text{bins} = n_\text{bins}^2$.  With this choice the one dimensional sample sizes are
\begin{equation}
N^{a}_\text{ev}=N^{b}_\text{ev} = \frac{2 N^{ab}_\text{ev}}{n_\text{bins}} \,.
\end{equation}
With this choice, we find that the mutual information that we calculate on Monte Carlo data is largely independent of the number of bins $n_\text{bins}$.  There do exist non-parametric methods for estimating mutual information that often have the advantage that a significantly smaller amount of data can be used.  We discuss one such algorithm in \App{app:nonpara_mi}.

\begin{figure}[t]
\includegraphics[width=7.4cm]{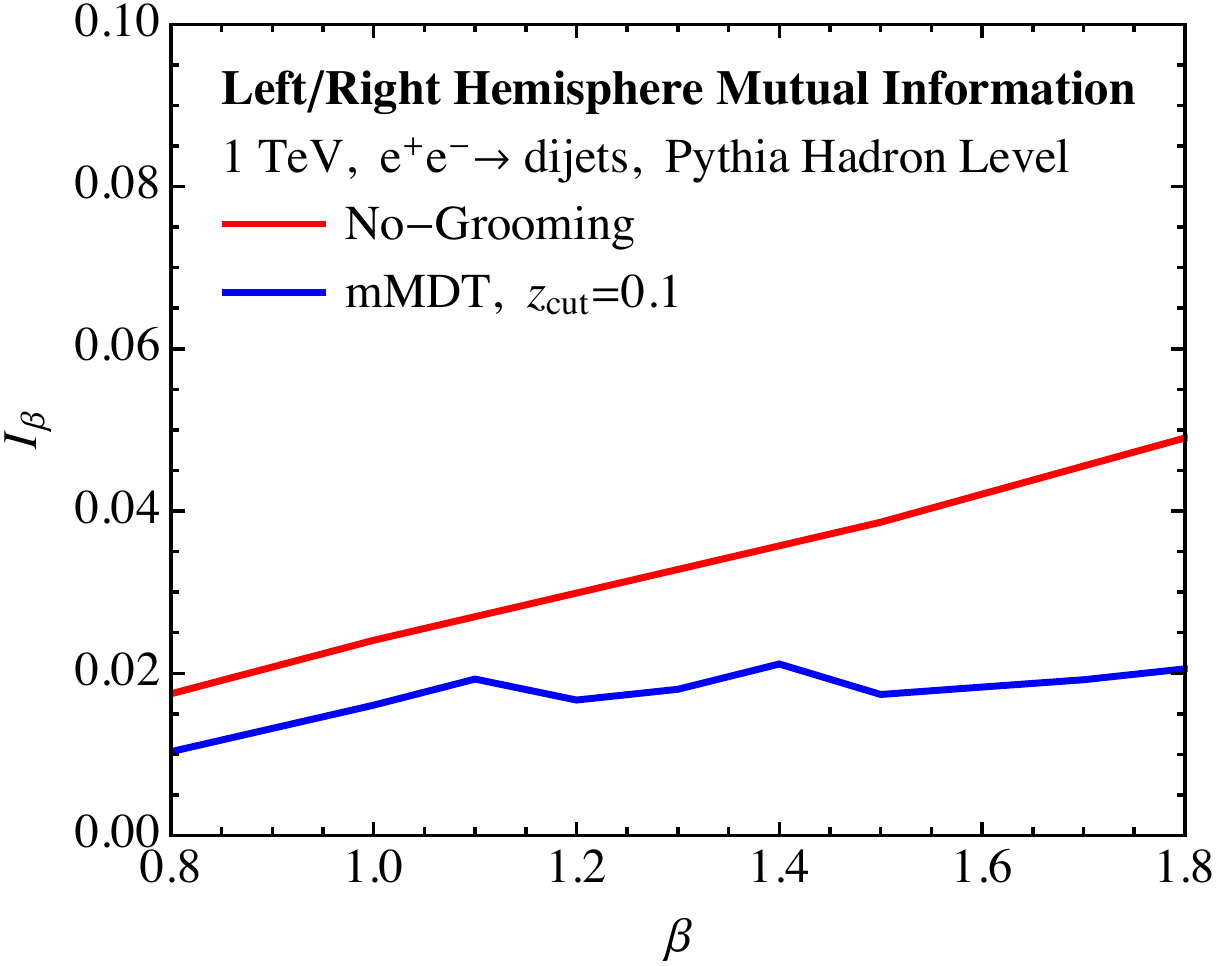} 
\caption{  The mutual information for  $e^+e^-\to $ dijets before and after the application of the modified mass drop (mMDT) grooming procedure.
}
\label{fig:mi_softdrop}
\end{figure}

We plot the extracted mutual information $I_\beta$ in both parton level and hadron level Monte Carlo in \Fig{fig:mi_MC}.  Both Monte Carlos exhibit the characteristic increase in $I_\beta$ as $\beta$ increases, indicative of non-global correlations.  The values of $I_\beta$ are nicely consistent with our calculations at both parton level (\Fig{fig:pert_mi}) and hadron level (\Fig{fig:pert_mi_np}), as well as with each other. We do not have a clear explanation for the slight discrepancy between the two generators observed at higher $\beta$ at parton level, as it is well within our perturbative uncertainties.  The Monte Carlo also manifests the expected increase of non-global correlations with the addition of hadronization.  

Non-global effects and NGLs in particular, strictly first occur at NLL order, and this observable demonstrates that formally leading logarithmic parton showers do include some amount of NGLs. This is perhaps not too surprising, because even in the Monte Carlos it is possible that an emission in the left hemisphere subsequently emits into the right hemisphere.  What may be surprising is that the strength of these correlations seems to be consistent with that calculated to NLL accuracy.  Because of the slightly different physics incorporated in the Monte Carlo predictions and the analytic calculations, we have chosen not to show them on the same plot. As non-global effects are probed in more detail, it will important to understand, preferably with a proof from the evolution kernels in the parton shower, to what extent parton shower Monte Carlos reproduce, for example, the BMS equation. It would also be interesting to improve the perturbative uncertainties in the analytic calculation of the mutual information by including higher order NGLs, for example using the formalisms proposed in \Refs{Caron-Huot:2015bja,Larkoski:2015zka,Becher:2015hka,Neill:2015nya}, allowing for a more precise comparison between analytic calculations and Monte Carlo programs.

To provide further support for our interpretation of the rise of $I_\beta$ with $\beta$ being due to non-global interactions of soft QCD particles, in \Fig{fig:mi_softdrop} we show a comparison of the mutual information on \pythia{} dijet events measured before and after the application of the modified mass drop (or equivalently, soft drop) grooming procedure \cite{Dasgupta:2013ihk,Dasgupta:2013via,Larkoski:2014wba} with fractional energy cut $z_{\text{cut}}=0.1$. In analytic calculations, this jet groomer has been shown to eliminate NGLs by removing wide angle soft radiation from the jet. This prediction is supported by \Fig{fig:mi_softdrop}, where the groomed $I_\beta$ is small relative to $I_\beta$ measured on the original event, and exhibits minimal dependence on $\beta$. This suggests that soft wide angle radiation is not responsible for the residual correlations. We have restricted the range of the plot to a single unit interval in $\beta$, as for angular exponents $\beta\gtrsim2$, the values of the energy correlation function after grooming become very small and our numerical methods become unstable. Because the characteristic increase of $I_\beta$ with $\beta$ is not present in the groomed events, NGLs have indeed been removed.

\section{Conclusions}\label{sec:conc}

Despite their importance for precision jet calculations, NGLs have never directly been observed.  In this paper, we introduced a novel procedure that allows us to isolate non-global correlations and their manifestation as NGLs.  The correlation between hemispheres in $e^+e^-$ collisions as quantified by the mutual information is calculable and measurable and we find good agreement between analytics and Monte Carlo simulation.  This correlation is also very sensitive to the value of the coupling $\alpha_s$, which can be brought under better theoretical control with higher-order calculations.  Non-perturbative contributions generically increase the correlation between the hemispheres, but are still subdominant effects to perturbative NGL correlations.  The mutual information $I_\beta$ is sensitive to physics formally beyond the accuracy of Monte Carlo generators, and so could be a powerful observable for tuning.  Additionally, while we focused on correlations in TeV collisions, this procedure could be applied for collisions at the $Z$ pole at LEP.  Perturbative non-global correlations would still exist, but non-perturbative correlations would be significantly larger.

While we have focused our discussion on the mutual in formation in the  context of $e^+e^-$ collisions it, or something similar, can in principle be measured at a hadron collider. A potentially clean measurement which is directly related to our discussion at $e^+e^-$ colliders is the decay of a color singlet initial state into hadrons. One could create such a sample at the LHC by identifying the hadronic decays of electroweak bosons.  With all particles from the decay, one can boost the system to its center of mass, identify the hemispheres, and measure the correlations.  It would also be interesting to use this observable to measure correlations between jet properties in $pp \to$ dijets, or more generally in QCD events. In particular, this procedure can be entirely data driven, and could be used as a powerful probe of correlations for measurements made in distinct regions of the detector. Experimentally, this would be significantly challenging, especially controlling effects from contamination.  Nevertheless, because it probes explicitly higher-order physics, measuring these non-global correlations could provide powerful insight into subleading QCD effects.

\begin{acknowledgments}
We thank Ben Nachman, Duff Neill, Iain Stewart and Jesse Thaler for helpful discussions and comments on the manuscript. This work is supported by the U.S. Department of Energy (DOE) under cooperative research agreements DE-FG02-05ER-41360, and DE-SC0011090. 
A.L.~is supported by the U.S. National Science Foundation, under grant PHY--1419008, the LHC Theory Initiative.  I.M.~is also supported by NSERC of Canada.  
\end{acknowledgments}

\appendix

\section{Global Logarithm Resummation}\label{app:resum}

In this appendix, we collect the expression of the resummed cross section to NLL order for global logarithms in the framework of SCET.  The first factorization theorem and resummation of the global logarithms of jet observables to NLL was presented in \Ref{Ellis:2010rwa}.  \Ref{Larkoski:2014uqa} extended that analysis to the resummation of recoil-free angularities \cite{Berger:2003iw,Almeida:2008yp} and energy correlation functions.  To NLL accuracy, the energy correlation functions were also resummed by the CAESAR collaboration \cite{Banfi:2004yd}.  The master formula for resummation of global logarithms of the hemisphere energy correlation functions is
\begin{widetext}
\begin{align}\label{eq:master_nll}
\sigma^{\text{G}}\left(\ecf{2,L}{\beta},\ecf{2,R}{\beta}\right) &= \exp\left[
K_H(\mu,\mu_H) + K_{J_R}(\mu,\mu_{J_R})+ K_{J_L}(\mu,\mu_{J_L})+ K_{S_R}(\mu,\mu_{S_R})+ K_{S_L}(\mu,\mu_{S_L})
\right] \\
&
\hspace{-1.5cm}
\times \exp\left[
\frac{2\gamma_E}{\beta}\left(
\omega_{J_R}(\mu,\mu_{J_R})+\omega_{J_L}(\mu,\mu_{J_L})
\right)+2\left(
\gamma_E+(\beta-1)\log 2
\right)\left(
\omega_{S_R}(\mu,\mu_{S_R})+\omega_{S_l}(\mu,\mu_{S_L})
\right)
\right] \left(
\frac{\mu_H^2}{4E_J^2}
\right)^{\omega_{H}(\mu,\mu_H)}\nonumber \\
&
\hspace{-1.5cm}
\times
\left(
\frac{\mu_{J_R}^2}{\left(\ecf{2,R}{\beta}\right)^{2/\beta}E_J^2}
\right)^{\omega_{J_R}(\mu,\mu_{J_R})}\left(
\frac{\mu_{J_L}^2}{\left(\ecf{2,L}{\beta}\right)^{2/\beta}E_J^2}
\right)^{\omega_{J_L}(\mu,\mu_{J_L})}\left(
\frac{\mu_{S_R}^2}{\left(\ecf{2,R}{\beta}\right)^{2}E_J^2}
\right)^{\omega_{S_R}(\mu,\mu_{S_R})}\left(
\frac{\mu_{S_L}^2}{\left(\ecf{2,L}{\beta}\right)^{2}E_J^2}
\right)^{\omega_{S_L}(\mu,\mu_{S_L})} \nonumber \\
&
\times\frac{1}{\Gamma\left(
1-\frac{2}{\beta}\omega_{J_r}(\mu,\mu_{J_R})-2\omega_{S_R}(\mu,\mu_{S_R})
\right)}\frac{1}{\Gamma\left(
1-\frac{2}{\beta}\omega_{J_L}(\mu,\mu_{J_L})-2\omega_{S_L}(\mu,\mu_{S_L})
\right)} \nonumber\,.
\end{align}
\end{widetext}
The functions $K_i(\mu,\mu_i)$ and $\omega_i(\mu,\mu_i)$ are defined as
\begin{widetext}
\begin{align}
K_i(\mu,\mu_i) &= C_i\frac{\Gamma_0}{2\beta_0^2}\left[
\frac{4\pi}{\alpha_s(\mu_i)}\left(
\log\, r+\frac{1}{r}-1
\right)+\left(
\frac{\Gamma_1}{\Gamma_0}-\frac{\beta_1}{\beta_0}
\right)\left(
r-1-\log\, r
\right)-\frac{\beta_1}{2\beta_0}\log^2 r
\right]-\frac{\gamma_0}{2\beta_0}\log \, r \, , \\
\omega_i(\mu,\mu_i) &=-C_i\frac{\Gamma_0}{2\beta_0}\left[
\log\, r+\frac{\alpha_s(\mu_i)}{4\pi}\left(
\frac{\Gamma_1}{\Gamma_0}-\frac{\beta_1}{\beta_0}
\right)(r-1)
\right]\,,
\nonumber
\end{align}
\end{widetext}
and the various QCD constants are
\begin{align}
\beta_0 &= \frac{11}{3}C_A - \frac{2}{3}n_f \,, \\
\beta_1 &= \frac{34}{3}C_A^2-\frac{10}{3} C_A n_f -2 C_F n_f\,,\\
\Gamma_0 &= 4\,,\\
\Gamma_1 &= 4 C_A\left(
\frac{67}{9}-\frac{\pi^2}{3}
\right)-\frac{40}{9}n_f\,,\\
r &= \frac{\alpha_s(\mu)}{\alpha_s(\mu_i)}\,.
\end{align}
The color factors and non-cusp anomalous dimensions appearing in the $K$ and $\omega$ functions are:
\begin{align}
C_H &= -2 C_F \,,\\
\gamma_H &= -12 C_F \,,\\
C_J &= \frac{\beta}{\beta-1}C_F\,,\\
\gamma_J &=6 C_F\,,\\
C_S &=\frac{1}{1-\beta}C_F \,,\\
\gamma_S &= 0\,.
\end{align}
For canonical resummation of the hemisphere observables to NLL, one sets the scales $\mu_i$ to their canonical values as determined by the ratios appearing in \Eq{eq:master_nll}.  By consistency of the factorization, the master formula is independent of the scale $\mu$.  By varying the scales $\mu_i$ in \Eq{eq:master_nll} by order-1 amounts, one can estimate the perturbative uncertainty by not resumming to higher accuracy.

\begin{figure*}[t]
\centering
\subfloat[]{\label{fig:npara1_v}
\includegraphics[width=7.5cm]{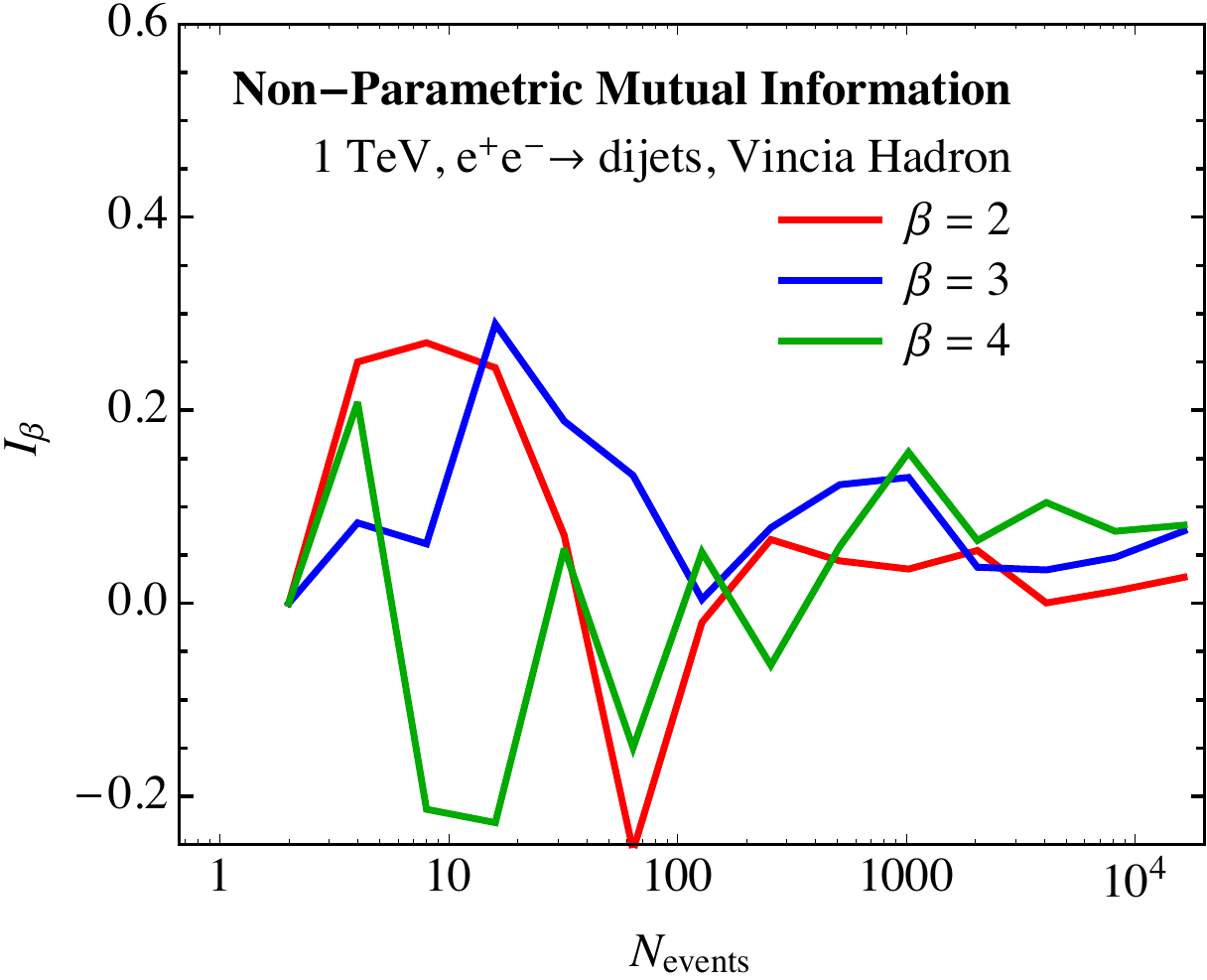}    
} \qquad\qquad
\subfloat[]{\label{fig:npara1_p}
\includegraphics[width=7.4cm]{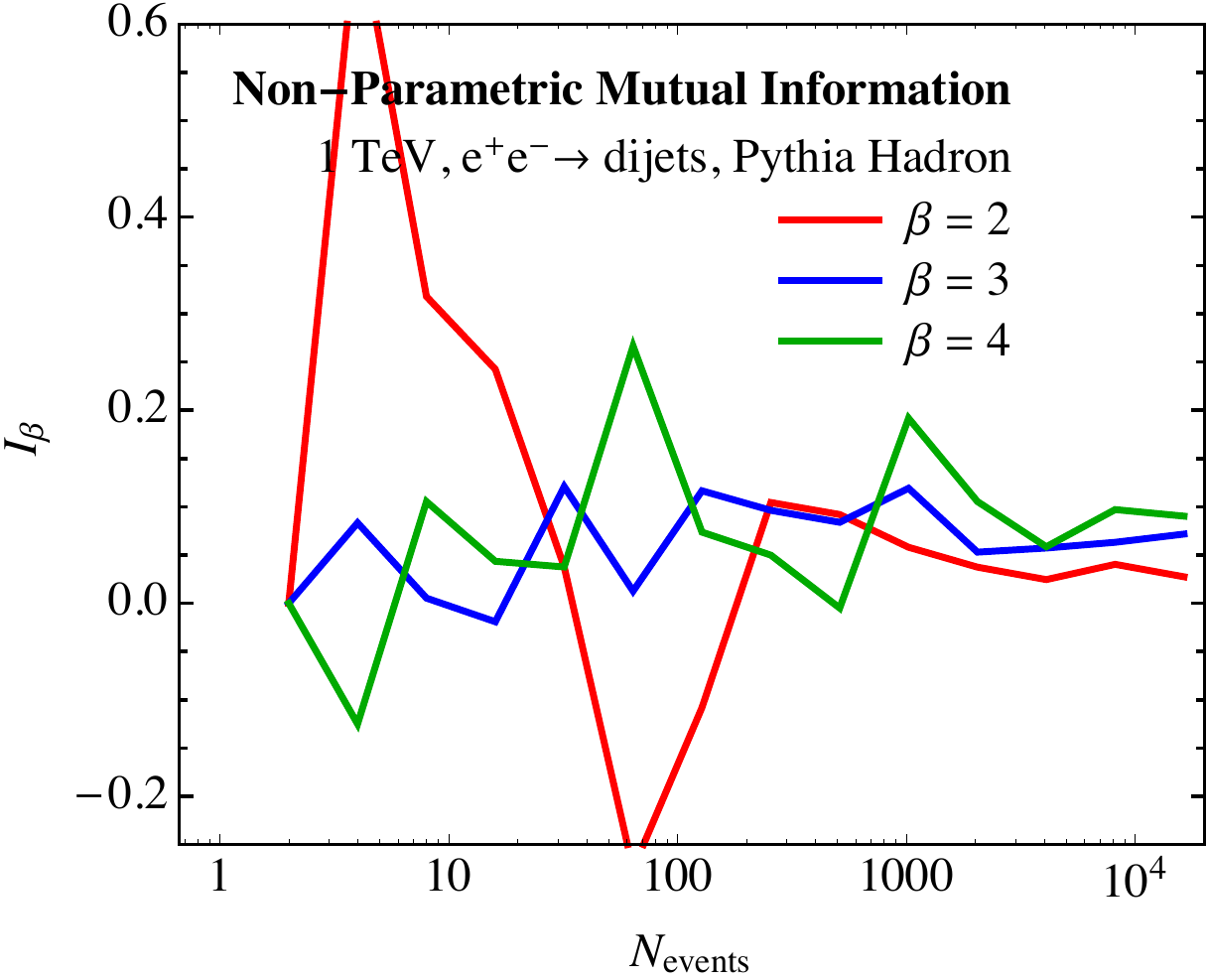} 
}\\
\caption{The non-parametric mutual information dependence on the number of events used in the sample.  Hadron-level \vincia{} is shown in (a) and \pythia{} in (b).
}
\label{fig:npara_1}
\end{figure*}

\begin{figure*}[t]
\centering
\subfloat[]{\label{fig:npara2_v}
\includegraphics[width=7.5cm]{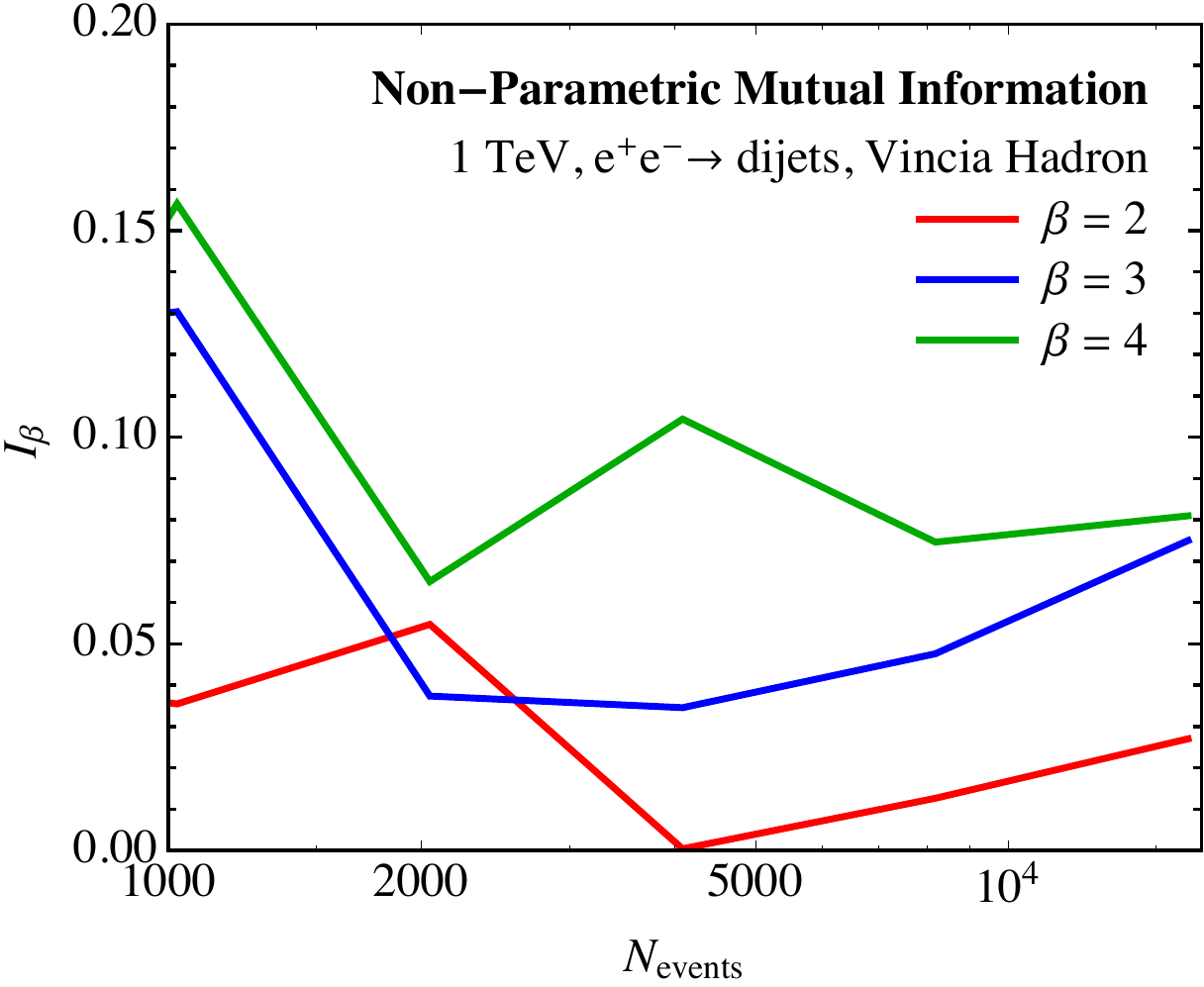}    
} \qquad\qquad
\subfloat[]{\label{fig:npara2_p}
\includegraphics[width=7.4cm]{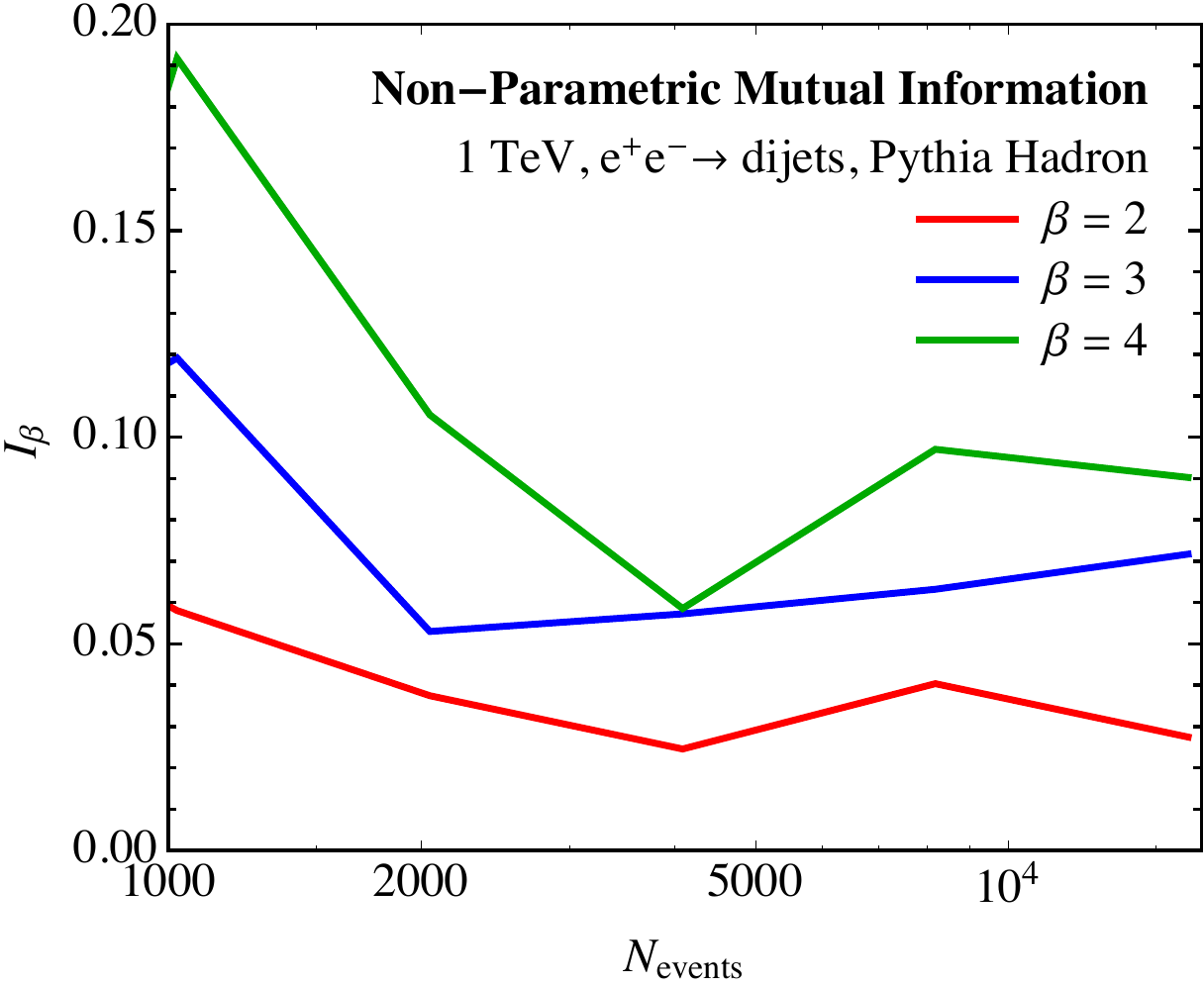} 
}\\
\caption{The non-parametric mutual information dependence on the number of events used in the sample, zoomed in to larger sample sizes.  Hadron-level \vincia{} is shown in (a) and \pythia{} in (b).
}
\label{fig:npara_2}
\end{figure*}

\section{Non-Parametric Algorithm for Mutual Information}\label{app:nonpara_mi}

In this appendix, we describe a non-parametric algorithm for calculating the mutual information on a finite data set, introduced in \Ref{2004PhRvE..69f6138K}.  We will not present a proof of the algorithm here, but just describe the method and demonstrate its performance on simulated Monte Carlo data.

Given a two-dimensional data set over an $N$ number of events,
\begin{equation}
Z=\{X,Y\}\,,
\end{equation}
 we define the distance measure between two events as
\begin{align}
||z-z'||=\max\left\{||x-x'||,||y-y'||\right\}\,,
\end{align}
where $||\ ||$ denotes any metric; for example, just the absolute value of the difference of measured values of the observables in the two events.  Here, $X$ or $Y$ denote the set of events and $x$ and $y$ denote the measured values of a particular event.  Then, given an integer $k$, for each event measurement $z_i$, one determines the minimal distance $\epsilon_i$ from $z_i$ in which $k$ other events are contained.  The estimate of the mutual information is then
\begin{align}
\hspace{-0.25cm}I(X,Y)=\psi(N)+\psi(k)-\left\langle\psi(n_x+1)+\psi(n_y+1)\right\rangle
\end{align}
where $\psi(x)$ is the digamma function.  $n_x(i)$ and $n_y(i)$ are the number of events which have measured values of $x$ and $y$, respectively, that are within a distance of $\epsilon_i$ of event $i$.  The angle brackets $\langle\rangle$ denote averaging over the ensemble.

We demonstrate in \Figs{fig:npara_1}{fig:npara_2} the effectiveness of this procedure for calculating the mutual information of the energy correlation functions of the two hemispheres from Monte Carlo simulation, as a function of the number of events $N$ used in the sample.  In these plots, we take $k=1$.  While it takes several thousand events for the fluctuations of the mutual information to stabilize, for $N\gtrsim 10^4$, the non-parametric mutual information does exhibit the expected ordering  as a function of angular exponent $\beta$.

\bibliography{mutinf_ngl}

\begin{thebibliography}{71}%
\makeatletter
\providecommand \@ifxundefined [1]{%
 \@ifx{#1\undefined}
}%
\providecommand \@ifnum [1]{%
 \ifnum #1\expandafter \@firstoftwo
 \else \expandafter \@secondoftwo
 \fi
}%
\providecommand \@ifx [1]{%
 \ifx #1\expandafter \@firstoftwo
 \else \expandafter \@secondoftwo
 \fi
}%
\providecommand \natexlab [1]{#1}%
\providecommand \enquote  [1]{``#1''}%
\providecommand \bibnamefont  [1]{#1}%
\providecommand \bibfnamefont [1]{#1}%
\providecommand \citenamefont [1]{#1}%
\providecommand \href@noop [0]{\@secondoftwo}%
\providecommand \href [0]{\begingroup \@sanitize@url \@href}%
\providecommand \@href[1]{\@@startlink{#1}\@@href}%
\providecommand \@@href[1]{\endgroup#1\@@endlink}%
\providecommand \@sanitize@url [0]{\catcode `\\12\catcode `\$12\catcode
  `\&12\catcode `\#12\catcode `\^12\catcode `\_12\catcode `\%12\relax}%
\providecommand \@@startlink[1]{}%
\providecommand \@@endlink[0]{}%
\providecommand \url  [0]{\begingroup\@sanitize@url \@url }%
\providecommand \@url [1]{\endgroup\@href {#1}{\urlprefix }}%
\providecommand \urlprefix  [0]{URL }%
\providecommand \Eprint [0]{\href }%
\providecommand \doibase [0]{http://dx.doi.org/}%
\providecommand \selectlanguage [0]{\@gobble}%
\providecommand \bibinfo  [0]{\@secondoftwo}%
\providecommand \bibfield  [0]{\@secondoftwo}%
\providecommand \translation [1]{[#1]}%
\providecommand \BibitemOpen [0]{}%
\providecommand \bibitemStop [0]{}%
\providecommand \bibitemNoStop [0]{.\EOS\space}%
\providecommand \EOS [0]{\spacefactor3000\relax}%
\providecommand \BibitemShut  [1]{\csname bibitem#1\endcsname}%
\let\auto@bib@innerbib\@empty
%</preamble>
\bibitem [{\citenamefont {Sterman}\ and\ \citenamefont
  {Weinberg}(1977)}]{Sterman:1977wj}%
  \BibitemOpen
  \bibfield  {author} {\bibinfo {author} {\bibfnamefont {G.~F.}\ \bibnamefont
  {Sterman}}\ and\ \bibinfo {author} {\bibfnamefont {S.}~\bibnamefont
  {Weinberg}},\ }\href {\doibase 10.1103/PhysRevLett.39.1436} {\bibfield
  {journal} {\bibinfo  {journal} {Phys. Rev. Lett.}\ }\textbf {\bibinfo
  {volume} {39}},\ \bibinfo {pages} {1436} (\bibinfo {year}
  {1977})}\BibitemShut {NoStop}%
%%CITATION = PRLTA,39,1436;%%
\bibitem [{\citenamefont {Dasgupta}\ and\ \citenamefont
  {Salam}(2001)}]{Dasgupta:2001sh}%
  \BibitemOpen
  \bibfield  {author} {\bibinfo {author} {\bibfnamefont {M.}~\bibnamefont
  {Dasgupta}}\ and\ \bibinfo {author} {\bibfnamefont {G.~P.}\ \bibnamefont
  {Salam}},\ }\href {\doibase 10.1016/S0370-2693(01)00725-0} {\bibfield
  {journal} {\bibinfo  {journal} {Phys. Lett.}\ }\textbf {\bibinfo {volume}
  {B512}},\ \bibinfo {pages} {323} (\bibinfo {year} {2001})},\ \Eprint
  {http://arxiv.org/abs/hep-ph/0104277} {arXiv:hep-ph/0104277 [hep-ph]}
  \BibitemShut {NoStop}%
%%CITATION = HEP-PH/0104277;%%
\bibitem [{\citenamefont {Kelley}\ \emph {et~al.}(2011)\citenamefont {Kelley},
  \citenamefont {Schwartz},\ and\ \citenamefont {Zhu}}]{Kelley:2011tj}%
  \BibitemOpen
  \bibfield  {author} {\bibinfo {author} {\bibfnamefont {R.}~\bibnamefont
  {Kelley}}, \bibinfo {author} {\bibfnamefont {M.~D.}\ \bibnamefont
  {Schwartz}}, \ and\ \bibinfo {author} {\bibfnamefont {H.~X.}\ \bibnamefont
  {Zhu}},\ }\href@noop {} {\  (\bibinfo {year} {2011})},\ \Eprint
  {http://arxiv.org/abs/1102.0561} {arXiv:1102.0561 [hep-ph]} \BibitemShut
  {NoStop}%
%%CITATION = ARXIV:1102.0561;%%
\bibitem [{\citenamefont {Chien}\ \emph {et~al.}(2013)\citenamefont {Chien},
  \citenamefont {Kelley}, \citenamefont {Schwartz},\ and\ \citenamefont
  {Zhu}}]{Chien:2012ur}%
  \BibitemOpen
  \bibfield  {author} {\bibinfo {author} {\bibfnamefont {Y.-T.}\ \bibnamefont
  {Chien}}, \bibinfo {author} {\bibfnamefont {R.}~\bibnamefont {Kelley}},
  \bibinfo {author} {\bibfnamefont {M.~D.}\ \bibnamefont {Schwartz}}, \ and\
  \bibinfo {author} {\bibfnamefont {H.~X.}\ \bibnamefont {Zhu}},\ }\href
  {\doibase 10.1103/PhysRevD.87.014010} {\bibfield  {journal} {\bibinfo
  {journal} {Phys. Rev.}\ }\textbf {\bibinfo {volume} {D87}},\ \bibinfo {pages}
  {014010} (\bibinfo {year} {2013})},\ \Eprint {http://arxiv.org/abs/1208.0010}
  {arXiv:1208.0010} \BibitemShut {NoStop}%
%%CITATION = ARXIV:1208.0010;%%
\bibitem [{\citenamefont {Dasgupta}\ \emph {et~al.}(2012)\citenamefont
  {Dasgupta}, \citenamefont {Khelifa-Kerfa}, \citenamefont {Marzani},\ and\
  \citenamefont {Spannowsky}}]{Dasgupta:2012hg}%
  \BibitemOpen
  \bibfield  {author} {\bibinfo {author} {\bibfnamefont {M.}~\bibnamefont
  {Dasgupta}}, \bibinfo {author} {\bibfnamefont {K.}~\bibnamefont
  {Khelifa-Kerfa}}, \bibinfo {author} {\bibfnamefont {S.}~\bibnamefont
  {Marzani}}, \ and\ \bibinfo {author} {\bibfnamefont {M.}~\bibnamefont
  {Spannowsky}},\ }\href {\doibase 10.1007/JHEP10(2012)126} {\bibfield
  {journal} {\bibinfo  {journal} {JHEP}\ }\textbf {\bibinfo {volume} {1210}},\
  \bibinfo {pages} {126} (\bibinfo {year} {2012})},\ \Eprint
  {http://arxiv.org/abs/1207.1640} {arXiv:1207.1640 [hep-ph]} \BibitemShut
  {NoStop}%
%%CITATION = ARXIV:1207.1640;%%
\bibitem [{\citenamefont {Jouttenus}\ \emph {et~al.}(2013)\citenamefont
  {Jouttenus}, \citenamefont {Stewart}, \citenamefont {Tackmann},\ and\
  \citenamefont {Waalewijn}}]{Jouttenus:2013hs}%
  \BibitemOpen
  \bibfield  {author} {\bibinfo {author} {\bibfnamefont {T.~T.}\ \bibnamefont
  {Jouttenus}}, \bibinfo {author} {\bibfnamefont {I.~W.}\ \bibnamefont
  {Stewart}}, \bibinfo {author} {\bibfnamefont {F.~J.}\ \bibnamefont
  {Tackmann}}, \ and\ \bibinfo {author} {\bibfnamefont {W.~J.}\ \bibnamefont
  {Waalewijn}},\ }\href {\doibase 10.1103/PhysRevD.88.054031} {\bibfield
  {journal} {\bibinfo  {journal} {Phys.Rev.}\ }\textbf {\bibinfo {volume}
  {D88}},\ \bibinfo {pages} {054031} (\bibinfo {year} {2013})},\ \Eprint
  {http://arxiv.org/abs/1302.0846} {arXiv:1302.0846 [hep-ph]} \BibitemShut
  {NoStop}%
%%CITATION = ARXIV:1302.0846;%%
\bibitem [{\citenamefont {Dasgupta}\ and\ \citenamefont
  {Salam}(2002{\natexlab{a}})}]{Dasgupta:2002bw}%
  \BibitemOpen
  \bibfield  {author} {\bibinfo {author} {\bibfnamefont {M.}~\bibnamefont
  {Dasgupta}}\ and\ \bibinfo {author} {\bibfnamefont {G.~P.}\ \bibnamefont
  {Salam}},\ }\href {\doibase 10.1088/1126-6708/2002/03/017} {\bibfield
  {journal} {\bibinfo  {journal} {JHEP}\ }\textbf {\bibinfo {volume} {0203}},\
  \bibinfo {pages} {017} (\bibinfo {year} {2002}{\natexlab{a}})},\ \Eprint
  {http://arxiv.org/abs/hep-ph/0203009} {arXiv:hep-ph/0203009 [hep-ph]}
  \BibitemShut {NoStop}%
%%CITATION = HEP-PH/0203009;%%
\bibitem [{\citenamefont {Dasgupta}\ and\ \citenamefont
  {Salam}(2002{\natexlab{b}})}]{Dasgupta:2002dc}%
  \BibitemOpen
  \bibfield  {author} {\bibinfo {author} {\bibfnamefont {M.}~\bibnamefont
  {Dasgupta}}\ and\ \bibinfo {author} {\bibfnamefont {G.~P.}\ \bibnamefont
  {Salam}},\ }\href {\doibase 10.1088/1126-6708/2002/08/032} {\bibfield
  {journal} {\bibinfo  {journal} {JHEP}\ }\textbf {\bibinfo {volume} {0208}},\
  \bibinfo {pages} {032} (\bibinfo {year} {2002}{\natexlab{b}})},\ \Eprint
  {http://arxiv.org/abs/hep-ph/0208073} {arXiv:hep-ph/0208073 [hep-ph]}
  \BibitemShut {NoStop}%
%%CITATION = HEP-PH/0208073;%%
\bibitem [{\citenamefont {Banfi}\ \emph {et~al.}(2002)\citenamefont {Banfi},
  \citenamefont {Marchesini},\ and\ \citenamefont {Smye}}]{Banfi:2002hw}%
  \BibitemOpen
  \bibfield  {author} {\bibinfo {author} {\bibfnamefont {A.}~\bibnamefont
  {Banfi}}, \bibinfo {author} {\bibfnamefont {G.}~\bibnamefont {Marchesini}}, \
  and\ \bibinfo {author} {\bibfnamefont {G.}~\bibnamefont {Smye}},\ }\href
  {\doibase 10.1088/1126-6708/2002/08/006} {\bibfield  {journal} {\bibinfo
  {journal} {JHEP}\ }\textbf {\bibinfo {volume} {0208}},\ \bibinfo {pages}
  {006} (\bibinfo {year} {2002})},\ \Eprint
  {http://arxiv.org/abs/hep-ph/0206076} {arXiv:hep-ph/0206076 [hep-ph]}
  \BibitemShut {NoStop}%
%%CITATION = HEP-PH/0206076;%%
\bibitem [{\citenamefont {Appleby}\ and\ \citenamefont
  {Seymour}(2002)}]{Appleby:2002ke}%
  \BibitemOpen
  \bibfield  {author} {\bibinfo {author} {\bibfnamefont {R.}~\bibnamefont
  {Appleby}}\ and\ \bibinfo {author} {\bibfnamefont {M.}~\bibnamefont
  {Seymour}},\ }\href {\doibase 10.1088/1126-6708/2002/12/063} {\bibfield
  {journal} {\bibinfo  {journal} {JHEP}\ }\textbf {\bibinfo {volume} {0212}},\
  \bibinfo {pages} {063} (\bibinfo {year} {2002})},\ \Eprint
  {http://arxiv.org/abs/hep-ph/0211426} {arXiv:hep-ph/0211426 [hep-ph]}
  \BibitemShut {NoStop}%
%%CITATION = HEP-PH/0211426;%%
\bibitem [{\citenamefont {Weigert}(2004)}]{Weigert:2003mm}%
  \BibitemOpen
  \bibfield  {author} {\bibinfo {author} {\bibfnamefont {H.}~\bibnamefont
  {Weigert}},\ }\href {\doibase 10.1016/j.nuclphysb.2004.03.002} {\bibfield
  {journal} {\bibinfo  {journal} {Nucl.Phys.}\ }\textbf {\bibinfo {volume}
  {B685}},\ \bibinfo {pages} {321} (\bibinfo {year} {2004})},\ \Eprint
  {http://arxiv.org/abs/hep-ph/0312050} {arXiv:hep-ph/0312050 [hep-ph]}
  \BibitemShut {NoStop}%
%%CITATION = HEP-PH/0312050;%%
\bibitem [{\citenamefont {Rubin}(2010)}]{Rubin:2010fc}%
  \BibitemOpen
  \bibfield  {author} {\bibinfo {author} {\bibfnamefont {M.}~\bibnamefont
  {Rubin}},\ }\href {\doibase 10.1007/JHEP05(2010)005} {\bibfield  {journal}
  {\bibinfo  {journal} {JHEP}\ }\textbf {\bibinfo {volume} {1005}},\ \bibinfo
  {pages} {005} (\bibinfo {year} {2010})},\ \Eprint
  {http://arxiv.org/abs/1002.4557} {arXiv:1002.4557 [hep-ph]} \BibitemShut
  {NoStop}%
%%CITATION = ARXIV:1002.4557;%%
\bibitem [{\citenamefont {Banfi}\ \emph {et~al.}(2010)\citenamefont {Banfi},
  \citenamefont {Dasgupta}, \citenamefont {Khelifa-Kerfa},\ and\ \citenamefont
  {Marzani}}]{Banfi:2010pa}%
  \BibitemOpen
  \bibfield  {author} {\bibinfo {author} {\bibfnamefont {A.}~\bibnamefont
  {Banfi}}, \bibinfo {author} {\bibfnamefont {M.}~\bibnamefont {Dasgupta}},
  \bibinfo {author} {\bibfnamefont {K.}~\bibnamefont {Khelifa-Kerfa}}, \ and\
  \bibinfo {author} {\bibfnamefont {S.}~\bibnamefont {Marzani}},\ }\href
  {\doibase 10.1007/JHEP08(2010)064} {\bibfield  {journal} {\bibinfo  {journal}
  {JHEP}\ }\textbf {\bibinfo {volume} {1008}},\ \bibinfo {pages} {064}
  (\bibinfo {year} {2010})},\ \Eprint {http://arxiv.org/abs/1004.3483}
  {arXiv:1004.3483 [hep-ph]} \BibitemShut {NoStop}%
%%CITATION = ARXIV:1004.3483;%%
\bibitem [{\citenamefont {Hornig}\ \emph {et~al.}(2011)\citenamefont {Hornig},
  \citenamefont {Lee}, \citenamefont {Stewart}, \citenamefont {Walsh},\ and\
  \citenamefont {Zuberi}}]{Hornig:2011iu}%
  \BibitemOpen
  \bibfield  {author} {\bibinfo {author} {\bibfnamefont {A.}~\bibnamefont
  {Hornig}}, \bibinfo {author} {\bibfnamefont {C.}~\bibnamefont {Lee}},
  \bibinfo {author} {\bibfnamefont {I.~W.}\ \bibnamefont {Stewart}}, \bibinfo
  {author} {\bibfnamefont {J.~R.}\ \bibnamefont {Walsh}}, \ and\ \bibinfo
  {author} {\bibfnamefont {S.}~\bibnamefont {Zuberi}},\ }\href {\doibase
  10.1007/JHEP08(2011)054} {\bibfield  {journal} {\bibinfo  {journal} {JHEP}\
  }\textbf {\bibinfo {volume} {1108}},\ \bibinfo {pages} {054} (\bibinfo {year}
  {2011})},\ \Eprint {http://arxiv.org/abs/1105.4628} {arXiv:1105.4628
  [hep-ph]} \BibitemShut {NoStop}%
%%CITATION = ARXIV:1105.4628;%%
\bibitem [{\citenamefont {Hornig}\ \emph {et~al.}(2012)\citenamefont {Hornig},
  \citenamefont {Lee}, \citenamefont {Walsh},\ and\ \citenamefont
  {Zuberi}}]{Hornig:2011tg}%
  \BibitemOpen
  \bibfield  {author} {\bibinfo {author} {\bibfnamefont {A.}~\bibnamefont
  {Hornig}}, \bibinfo {author} {\bibfnamefont {C.}~\bibnamefont {Lee}},
  \bibinfo {author} {\bibfnamefont {J.~R.}\ \bibnamefont {Walsh}}, \ and\
  \bibinfo {author} {\bibfnamefont {S.}~\bibnamefont {Zuberi}},\ }\href
  {\doibase 10.1007/JHEP01(2012)149} {\bibfield  {journal} {\bibinfo  {journal}
  {JHEP}\ }\textbf {\bibinfo {volume} {1201}},\ \bibinfo {pages} {149}
  (\bibinfo {year} {2012})},\ \Eprint {http://arxiv.org/abs/1110.0004}
  {arXiv:1110.0004 [hep-ph]} \BibitemShut {NoStop}%
%%CITATION = ARXIV:1110.0004;%%
\bibitem [{\citenamefont {Kelley}\ \emph
  {et~al.}(2012{\natexlab{a}})\citenamefont {Kelley}, \citenamefont {Schwartz},
  \citenamefont {Schabinger},\ and\ \citenamefont {Zhu}}]{Kelley:2011aa}%
  \BibitemOpen
  \bibfield  {author} {\bibinfo {author} {\bibfnamefont {R.}~\bibnamefont
  {Kelley}}, \bibinfo {author} {\bibfnamefont {M.~D.}\ \bibnamefont
  {Schwartz}}, \bibinfo {author} {\bibfnamefont {R.~M.}\ \bibnamefont
  {Schabinger}}, \ and\ \bibinfo {author} {\bibfnamefont {H.~X.}\ \bibnamefont
  {Zhu}},\ }\href {\doibase 10.1103/PhysRevD.86.054017} {\bibfield  {journal}
  {\bibinfo  {journal} {Phys.Rev.}\ }\textbf {\bibinfo {volume} {D86}},\
  \bibinfo {pages} {054017} (\bibinfo {year} {2012}{\natexlab{a}})},\ \Eprint
  {http://arxiv.org/abs/1112.3343} {arXiv:1112.3343 [hep-ph]} \BibitemShut
  {NoStop}%
%%CITATION = ARXIV:1112.3343;%%
\bibitem [{\citenamefont {Kelley}\ \emph
  {et~al.}(2012{\natexlab{b}})\citenamefont {Kelley}, \citenamefont {Walsh},\
  and\ \citenamefont {Zuberi}}]{Kelley:2012kj}%
  \BibitemOpen
  \bibfield  {author} {\bibinfo {author} {\bibfnamefont {R.}~\bibnamefont
  {Kelley}}, \bibinfo {author} {\bibfnamefont {J.~R.}\ \bibnamefont {Walsh}}, \
  and\ \bibinfo {author} {\bibfnamefont {S.}~\bibnamefont {Zuberi}},\ }\href
  {\doibase 10.1007/JHEP09(2012)117} {\bibfield  {journal} {\bibinfo  {journal}
  {JHEP}\ }\textbf {\bibinfo {volume} {1209}},\ \bibinfo {pages} {117}
  (\bibinfo {year} {2012}{\natexlab{b}})},\ \Eprint
  {http://arxiv.org/abs/1202.2361} {arXiv:1202.2361 [hep-ph]} \BibitemShut
  {NoStop}%
%%CITATION = ARXIV:1202.2361;%%
\bibitem [{\citenamefont {Hatta}\ and\ \citenamefont
  {Ueda}(2013)}]{Hatta:2013iba}%
  \BibitemOpen
  \bibfield  {author} {\bibinfo {author} {\bibfnamefont {Y.}~\bibnamefont
  {Hatta}}\ and\ \bibinfo {author} {\bibfnamefont {T.}~\bibnamefont {Ueda}},\
  }\href {\doibase 10.1016/j.nuclphysb.2013.06.021} {\bibfield  {journal}
  {\bibinfo  {journal} {Nucl.Phys.}\ }\textbf {\bibinfo {volume} {B874}},\
  \bibinfo {pages} {808} (\bibinfo {year} {2013})},\ \Eprint
  {http://arxiv.org/abs/1304.6930} {arXiv:1304.6930 [hep-ph]} \BibitemShut
  {NoStop}%
%%CITATION = ARXIV:1304.6930;%%
\bibitem [{\citenamefont {Schwartz}\ and\ \citenamefont
  {Zhu}(2014)}]{Schwartz:2014wha}%
  \BibitemOpen
  \bibfield  {author} {\bibinfo {author} {\bibfnamefont {M.~D.}\ \bibnamefont
  {Schwartz}}\ and\ \bibinfo {author} {\bibfnamefont {H.~X.}\ \bibnamefont
  {Zhu}},\ }\href {\doibase 10.1103/PhysRevD.90.065004} {\bibfield  {journal}
  {\bibinfo  {journal} {Phys.Rev.}\ }\textbf {\bibinfo {volume} {D90}},\
  \bibinfo {pages} {065004} (\bibinfo {year} {2014})},\ \Eprint
  {http://arxiv.org/abs/1403.4949} {arXiv:1403.4949 [hep-ph]} \BibitemShut
  {NoStop}%
%%CITATION = ARXIV:1403.4949;%%
\bibitem [{\citenamefont {Khelifa-Kerfa}\ and\ \citenamefont
  {Delenda}(2015)}]{Khelifa-Kerfa:2015mma}%
  \BibitemOpen
  \bibfield  {author} {\bibinfo {author} {\bibfnamefont {K.}~\bibnamefont
  {Khelifa-Kerfa}}\ and\ \bibinfo {author} {\bibfnamefont {Y.}~\bibnamefont
  {Delenda}},\ }\href@noop {} {\  (\bibinfo {year} {2015})},\ \Eprint
  {http://arxiv.org/abs/1501.00475} {arXiv:1501.00475 [hep-ph]} \BibitemShut
  {NoStop}%
%%CITATION = ARXIV:1501.00475;%%
\bibitem [{\citenamefont {Caron-Huot}(2015)}]{Caron-Huot:2015bja}%
  \BibitemOpen
  \bibfield  {author} {\bibinfo {author} {\bibfnamefont {S.}~\bibnamefont
  {Caron-Huot}},\ }\href@noop {} {\  (\bibinfo {year} {2015})},\ \Eprint
  {http://arxiv.org/abs/1501.03754} {arXiv:1501.03754 [hep-ph]} \BibitemShut
  {NoStop}%
%%CITATION = ARXIV:1501.03754;%%
\bibitem [{\citenamefont {Larkoski}\ \emph {et~al.}(2015)\citenamefont
  {Larkoski}, \citenamefont {Moult},\ and\ \citenamefont
  {Neill}}]{Larkoski:2015zka}%
  \BibitemOpen
  \bibfield  {author} {\bibinfo {author} {\bibfnamefont {A.~J.}\ \bibnamefont
  {Larkoski}}, \bibinfo {author} {\bibfnamefont {I.}~\bibnamefont {Moult}}, \
  and\ \bibinfo {author} {\bibfnamefont {D.}~\bibnamefont {Neill}},\ }\href
  {\doibase 10.1007/JHEP09(2015)143} {\bibfield  {journal} {\bibinfo  {journal}
  {JHEP}\ }\textbf {\bibinfo {volume} {09}},\ \bibinfo {pages} {143} (\bibinfo
  {year} {2015})},\ \Eprint {http://arxiv.org/abs/1501.04596} {arXiv:1501.04596
  [hep-ph]} \BibitemShut {NoStop}%
%%CITATION = ARXIV:1501.04596;%%
\bibitem [{\citenamefont {Hagiwara}\ \emph {et~al.}(2015)\citenamefont
  {Hagiwara}, \citenamefont {Hatta},\ and\ \citenamefont
  {Ueda}}]{Hagiwara:2015bia}%
  \BibitemOpen
  \bibfield  {author} {\bibinfo {author} {\bibfnamefont {Y.}~\bibnamefont
  {Hagiwara}}, \bibinfo {author} {\bibfnamefont {Y.}~\bibnamefont {Hatta}}, \
  and\ \bibinfo {author} {\bibfnamefont {T.}~\bibnamefont {Ueda}},\ }\href@noop
  {} {\  (\bibinfo {year} {2015})},\ \Eprint {http://arxiv.org/abs/1507.07641}
  {arXiv:1507.07641 [hep-ph]} \BibitemShut {NoStop}%
%%CITATION = ARXIV:1507.07641;%%
\bibitem [{\citenamefont {Becher}\ \emph {et~al.}(2015)\citenamefont {Becher},
  \citenamefont {Neubert}, \citenamefont {Rothen},\ and\ \citenamefont
  {Shao}}]{Becher:2015hka}%
  \BibitemOpen
  \bibfield  {author} {\bibinfo {author} {\bibfnamefont {T.}~\bibnamefont
  {Becher}}, \bibinfo {author} {\bibfnamefont {M.}~\bibnamefont {Neubert}},
  \bibinfo {author} {\bibfnamefont {L.}~\bibnamefont {Rothen}}, \ and\ \bibinfo
  {author} {\bibfnamefont {D.~Y.}\ \bibnamefont {Shao}},\ }\href@noop {} {\
  (\bibinfo {year} {2015})},\ \Eprint {http://arxiv.org/abs/1508.06645}
  {arXiv:1508.06645 [hep-ph]} \BibitemShut {NoStop}%
%%CITATION = ARXIV:1508.06645;%%
\bibitem [{\citenamefont {Neill}(2015)}]{Neill:2015nya}%
  \BibitemOpen
  \bibfield  {author} {\bibinfo {author} {\bibfnamefont {D.}~\bibnamefont
  {Neill}},\ }\href@noop {} {\  (\bibinfo {year} {2015})},\ \Eprint
  {http://arxiv.org/abs/1508.07568} {arXiv:1508.07568 [hep-ph]} \BibitemShut
  {NoStop}%
%%CITATION = ARXIV:1508.07568;%%
\bibitem [{\citenamefont {Dasgupta}\ \emph
  {et~al.}(2013{\natexlab{a}})\citenamefont {Dasgupta}, \citenamefont
  {Fregoso}, \citenamefont {Marzani},\ and\ \citenamefont
  {Salam}}]{Dasgupta:2013ihk}%
  \BibitemOpen
  \bibfield  {author} {\bibinfo {author} {\bibfnamefont {M.}~\bibnamefont
  {Dasgupta}}, \bibinfo {author} {\bibfnamefont {A.}~\bibnamefont {Fregoso}},
  \bibinfo {author} {\bibfnamefont {S.}~\bibnamefont {Marzani}}, \ and\
  \bibinfo {author} {\bibfnamefont {G.~P.}\ \bibnamefont {Salam}},\ }\href
  {\doibase 10.1007/JHEP09(2013)029} {\bibfield  {journal} {\bibinfo  {journal}
  {JHEP}\ }\textbf {\bibinfo {volume} {1309}},\ \bibinfo {pages} {029}
  (\bibinfo {year} {2013}{\natexlab{a}})},\ \Eprint
  {http://arxiv.org/abs/1307.0007} {arXiv:1307.0007 [hep-ph]} \BibitemShut
  {NoStop}%
%%CITATION = ARXIV:1307.0007;%%
\bibitem [{\citenamefont {Dasgupta}\ \emph
  {et~al.}(2013{\natexlab{b}})\citenamefont {Dasgupta}, \citenamefont
  {Fregoso}, \citenamefont {Marzani},\ and\ \citenamefont
  {Powling}}]{Dasgupta:2013via}%
  \BibitemOpen
  \bibfield  {author} {\bibinfo {author} {\bibfnamefont {M.}~\bibnamefont
  {Dasgupta}}, \bibinfo {author} {\bibfnamefont {A.}~\bibnamefont {Fregoso}},
  \bibinfo {author} {\bibfnamefont {S.}~\bibnamefont {Marzani}}, \ and\
  \bibinfo {author} {\bibfnamefont {A.}~\bibnamefont {Powling}},\ }\href
  {\doibase 10.1140/epjc/s10052-013-2623-3} {\bibfield  {journal} {\bibinfo
  {journal} {Eur.Phys.J.}\ }\textbf {\bibinfo {volume} {C73}},\ \bibinfo
  {pages} {2623} (\bibinfo {year} {2013}{\natexlab{b}})},\ \Eprint
  {http://arxiv.org/abs/1307.0013} {arXiv:1307.0013 [hep-ph]} \BibitemShut
  {NoStop}%
%%CITATION = ARXIV:1307.0013;%%
\bibitem [{\citenamefont {Larkoski}\ \emph
  {et~al.}(2014{\natexlab{a}})\citenamefont {Larkoski}, \citenamefont
  {Marzani}, \citenamefont {Soyez},\ and\ \citenamefont
  {Thaler}}]{Larkoski:2014wba}%
  \BibitemOpen
  \bibfield  {author} {\bibinfo {author} {\bibfnamefont {A.~J.}\ \bibnamefont
  {Larkoski}}, \bibinfo {author} {\bibfnamefont {S.}~\bibnamefont {Marzani}},
  \bibinfo {author} {\bibfnamefont {G.}~\bibnamefont {Soyez}}, \ and\ \bibinfo
  {author} {\bibfnamefont {J.}~\bibnamefont {Thaler}},\ }\href {\doibase
  10.1007/JHEP05(2014)146} {\bibfield  {journal} {\bibinfo  {journal} {JHEP}\
  }\textbf {\bibinfo {volume} {1405}},\ \bibinfo {pages} {146} (\bibinfo {year}
  {2014}{\natexlab{a}})},\ \Eprint {http://arxiv.org/abs/1402.2657}
  {arXiv:1402.2657 [hep-ph]} \BibitemShut {NoStop}%
%%CITATION = ARXIV:1402.2657;%%
\bibitem [{\citenamefont {Banfi}\ \emph {et~al.}(2005)\citenamefont {Banfi},
  \citenamefont {Salam},\ and\ \citenamefont {Zanderighi}}]{Banfi:2004yd}%
  \BibitemOpen
  \bibfield  {author} {\bibinfo {author} {\bibfnamefont {A.}~\bibnamefont
  {Banfi}}, \bibinfo {author} {\bibfnamefont {G.~P.}\ \bibnamefont {Salam}}, \
  and\ \bibinfo {author} {\bibfnamefont {G.}~\bibnamefont {Zanderighi}},\
  }\href {\doibase 10.1088/1126-6708/2005/03/073} {\bibfield  {journal}
  {\bibinfo  {journal} {JHEP}\ }\textbf {\bibinfo {volume} {0503}},\ \bibinfo
  {pages} {073} (\bibinfo {year} {2005})},\ \Eprint
  {http://arxiv.org/abs/hep-ph/0407286} {arXiv:hep-ph/0407286 [hep-ph]}
  \BibitemShut {NoStop}%
%%CITATION = HEP-PH/0407286;%%
\bibitem [{\citenamefont {Salam}()}]{Salambroadening}%
  \BibitemOpen
  \bibfield  {author} {\bibinfo {author} {\bibfnamefont {G.}~\bibnamefont
  {Salam}},\ }\href@noop {} {\bibinfo  {journal} {Unpublished}\ }\BibitemShut
  {NoStop}%
%%CITATION = ARXIV:1307.1699;%%
\bibitem [{\citenamefont {Larkoski}\ \emph
  {et~al.}(2014{\natexlab{b}})\citenamefont {Larkoski}, \citenamefont {Neill},\
  and\ \citenamefont {Thaler}}]{Larkoski:2014uqa}%
  \BibitemOpen
\bibfield  {journal} {  }\bibfield  {author} {\bibinfo {author} {\bibfnamefont
  {A.~J.}\ \bibnamefont {Larkoski}}, \bibinfo {author} {\bibfnamefont
  {D.}~\bibnamefont {Neill}}, \ and\ \bibinfo {author} {\bibfnamefont
  {J.}~\bibnamefont {Thaler}},\ }\href {\doibase 10.1007/JHEP04(2014)017}
  {\bibfield  {journal} {\bibinfo  {journal} {JHEP}\ }\textbf {\bibinfo
  {volume} {1404}},\ \bibinfo {pages} {017} (\bibinfo {year}
  {2014}{\natexlab{b}})},\ \Eprint {http://arxiv.org/abs/1401.2158}
  {arXiv:1401.2158 [hep-ph]} \BibitemShut {NoStop}%
%%CITATION = ARXIV:1401.2158;%%
\bibitem [{\citenamefont {Jankowiak}\ and\ \citenamefont
  {Larkoski}(2011)}]{Jankowiak:2011qa}%
  \BibitemOpen
  \bibfield  {author} {\bibinfo {author} {\bibfnamefont {M.}~\bibnamefont
  {Jankowiak}}\ and\ \bibinfo {author} {\bibfnamefont {A.~J.}\ \bibnamefont
  {Larkoski}},\ }\href {\doibase 10.1007/JHEP06(2011)057} {\bibfield  {journal}
  {\bibinfo  {journal} {JHEP}\ }\textbf {\bibinfo {volume} {1106}},\ \bibinfo
  {pages} {057} (\bibinfo {year} {2011})},\ \Eprint
  {http://arxiv.org/abs/1104.1646} {arXiv:1104.1646 [hep-ph]} \BibitemShut
  {NoStop}%
%%CITATION = ARXIV:1104.1646;%%
\bibitem [{\citenamefont {Gallicchio}\ and\ \citenamefont
  {Schwartz}(2013)}]{Gallicchio:2012ez}%
  \BibitemOpen
  \bibfield  {author} {\bibinfo {author} {\bibfnamefont {J.}~\bibnamefont
  {Gallicchio}}\ and\ \bibinfo {author} {\bibfnamefont {M.~D.}\ \bibnamefont
  {Schwartz}},\ }\href {\doibase 10.1007/JHEP04(2013)090} {\bibfield  {journal}
  {\bibinfo  {journal} {JHEP}\ }\textbf {\bibinfo {volume} {1304}},\ \bibinfo
  {pages} {090} (\bibinfo {year} {2013})},\ \Eprint
  {http://arxiv.org/abs/1211.7038} {arXiv:1211.7038 [hep-ph]} \BibitemShut
  {NoStop}%
%%CITATION = ARXIV:1211.7038;%%
\bibitem [{\citenamefont {Larkoski}\ \emph
  {et~al.}(2013{\natexlab{a}})\citenamefont {Larkoski}, \citenamefont {Salam},\
  and\ \citenamefont {Thaler}}]{Larkoski:2013eya}%
  \BibitemOpen
  \bibfield  {author} {\bibinfo {author} {\bibfnamefont {A.~J.}\ \bibnamefont
  {Larkoski}}, \bibinfo {author} {\bibfnamefont {G.~P.}\ \bibnamefont {Salam}},
  \ and\ \bibinfo {author} {\bibfnamefont {J.}~\bibnamefont {Thaler}},\ }\href
  {\doibase 10.1007/JHEP06(2013)108} {\bibfield  {journal} {\bibinfo  {journal}
  {JHEP}\ }\textbf {\bibinfo {volume} {1306}},\ \bibinfo {pages} {108}
  (\bibinfo {year} {2013}{\natexlab{a}})},\ \Eprint
  {http://arxiv.org/abs/1305.0007} {arXiv:1305.0007 [hep-ph]} \BibitemShut
  {NoStop}%
%%CITATION = ARXIV:1305.0007;%%
\bibitem [{\citenamefont {Larkoski}\ \emph
  {et~al.}(2014{\natexlab{c}})\citenamefont {Larkoski}, \citenamefont
  {Thaler},\ and\ \citenamefont {Waalewijn}}]{Larkoski:2014pca}%
  \BibitemOpen
  \bibfield  {author} {\bibinfo {author} {\bibfnamefont {A.~J.}\ \bibnamefont
  {Larkoski}}, \bibinfo {author} {\bibfnamefont {J.}~\bibnamefont {Thaler}}, \
  and\ \bibinfo {author} {\bibfnamefont {W.~J.}\ \bibnamefont {Waalewijn}},\
  }\href {\doibase 10.1007/JHEP11(2014)129} {\bibfield  {journal} {\bibinfo
  {journal} {JHEP}\ }\textbf {\bibinfo {volume} {1411}},\ \bibinfo {pages}
  {129} (\bibinfo {year} {2014}{\natexlab{c}})},\ \Eprint
  {http://arxiv.org/abs/1408.3122} {arXiv:1408.3122 [hep-ph]} \BibitemShut
  {NoStop}%
%%CITATION = ARXIV:1408.3122;%%
\bibitem [{\citenamefont {Carruthers}\ and\ \citenamefont
  {Shih}(1989)}]{Carruthers:1989gu}%
  \BibitemOpen
  \bibfield  {author} {\bibinfo {author} {\bibfnamefont {P.}~\bibnamefont
  {Carruthers}}\ and\ \bibinfo {author} {\bibfnamefont {C.~C.}\ \bibnamefont
  {Shih}},\ }\href {\doibase 10.1103/PhysRevLett.62.2073} {\bibfield  {journal}
  {\bibinfo  {journal} {Phys. Rev. Lett.}\ }\textbf {\bibinfo {volume} {62}},\
  \bibinfo {pages} {2073} (\bibinfo {year} {1989})}\BibitemShut {NoStop}%
%%CITATION = PRLTA,62,2073;%%
\bibitem [{\citenamefont {Narsky}\ and\ \citenamefont
  {Porter}(2014)}]{Narsky:2014fya}%
  \BibitemOpen
  \bibfield  {author} {\bibinfo {author} {\bibfnamefont {I.}~\bibnamefont
  {Narsky}}\ and\ \bibinfo {author} {\bibfnamefont {F.~C.}\ \bibnamefont
  {Porter}},\ }\href
  {http://www.wiley-vch.de/publish/dt/books/ISBN3-527-41086-4} {\emph {\bibinfo
  {title} {{Statistical analysis techniques in particle physics}}}}\ (\bibinfo
  {publisher} {Wiley-VCH},\ \bibinfo {address} {Weinheim, Germany},\ \bibinfo
  {year} {2014})\BibitemShut {NoStop}%
%%CITATION = INSPIRE-1289761;%%
\bibitem [{\citenamefont {Bell}(1964)}]{Bell:1964kc}%
  \BibitemOpen
  \bibfield  {author} {\bibinfo {author} {\bibfnamefont {J.~S.}\ \bibnamefont
  {Bell}},\ }\href@noop {} {\bibfield  {journal} {\bibinfo  {journal}
  {Physics}\ }\textbf {\bibinfo {volume} {1}},\ \bibinfo {pages} {195}
  (\bibinfo {year} {1964})}\BibitemShut {NoStop}%
%%CITATION = PYCSA,1,195;%%
\bibitem [{\citenamefont {Maldacena}(2015)}]{Maldacena:2015bha}%
  \BibitemOpen
  \bibfield  {author} {\bibinfo {author} {\bibfnamefont {J.}~\bibnamefont
  {Maldacena}},\ }\href@noop {} {\  (\bibinfo {year} {2015})},\ \Eprint
  {http://arxiv.org/abs/1508.01082} {arXiv:1508.01082 [hep-th]} \BibitemShut
  {NoStop}%
%%CITATION = ARXIV:1508.01082;%%
\bibitem [{\citenamefont {Mangano}\ and\ \citenamefont
  {Parke}(1991)}]{Mangano:1990by}%
  \BibitemOpen
  \bibfield  {author} {\bibinfo {author} {\bibfnamefont {M.~L.}\ \bibnamefont
  {Mangano}}\ and\ \bibinfo {author} {\bibfnamefont {S.~J.}\ \bibnamefont
  {Parke}},\ }\href {\doibase 10.1016/0370-1573(91)90091-Y} {\bibfield
  {journal} {\bibinfo  {journal} {Phys. Rept.}\ }\textbf {\bibinfo {volume}
  {200}},\ \bibinfo {pages} {301} (\bibinfo {year} {1991})},\ \Eprint
  {http://arxiv.org/abs/hep-th/0509223} {arXiv:hep-th/0509223 [hep-th]}
  \BibitemShut {NoStop}%
%%CITATION = HEP-TH/0509223;%%
\bibitem [{\citenamefont {Dixon}(1996)}]{Dixon:1996wi}%
  \BibitemOpen
  \bibfield  {author} {\bibinfo {author} {\bibfnamefont {L.~J.}\ \bibnamefont
  {Dixon}},\ }in\ \href
  {http://www-public.slac.stanford.edu/sciDoc/docMeta.aspx?slacPubNumber=SLAC-PUB-7106}
  {\emph {\bibinfo {booktitle} {{QCD and beyond. Proceedings, Theoretical
  Advanced Study Institute in Elementary Particle Physics, TASI-95, Boulder,
  USA, June 4-30, 1995}}}}\ (\bibinfo {year} {1996})\ \Eprint
  {http://arxiv.org/abs/hep-ph/9601359} {arXiv:hep-ph/9601359 [hep-ph]}
  \BibitemShut {NoStop}%
%%CITATION = HEP-PH/9601359;%%
\bibitem [{\citenamefont {Banfi}\ and\ \citenamefont
  {Dasgupta}(2005)}]{Banfi:2005gj}%
  \BibitemOpen
  \bibfield  {author} {\bibinfo {author} {\bibfnamefont {A.}~\bibnamefont
  {Banfi}}\ and\ \bibinfo {author} {\bibfnamefont {M.}~\bibnamefont
  {Dasgupta}},\ }\href {\doibase 10.1016/j.physletb.2005.08.125} {\bibfield
  {journal} {\bibinfo  {journal} {Phys. Lett.}\ }\textbf {\bibinfo {volume}
  {B628}},\ \bibinfo {pages} {49} (\bibinfo {year} {2005})},\ \Eprint
  {http://arxiv.org/abs/hep-ph/0508159} {arXiv:hep-ph/0508159 [hep-ph]}
  \BibitemShut {NoStop}%
%%CITATION = HEP-PH/0508159;%%
\bibitem [{\citenamefont {Delenda}\ \emph {et~al.}(2006)\citenamefont
  {Delenda}, \citenamefont {Appleby}, \citenamefont {Dasgupta},\ and\
  \citenamefont {Banfi}}]{Delenda:2006nf}%
  \BibitemOpen
  \bibfield  {author} {\bibinfo {author} {\bibfnamefont {Y.}~\bibnamefont
  {Delenda}}, \bibinfo {author} {\bibfnamefont {R.}~\bibnamefont {Appleby}},
  \bibinfo {author} {\bibfnamefont {M.}~\bibnamefont {Dasgupta}}, \ and\
  \bibinfo {author} {\bibfnamefont {A.}~\bibnamefont {Banfi}},\ }\href
  {\doibase 10.1088/1126-6708/2006/12/044} {\bibfield  {journal} {\bibinfo
  {journal} {JHEP}\ }\textbf {\bibinfo {volume} {12}},\ \bibinfo {pages} {044}
  (\bibinfo {year} {2006})},\ \Eprint {http://arxiv.org/abs/hep-ph/0610242}
  {arXiv:hep-ph/0610242 [hep-ph]} \BibitemShut {NoStop}%
%%CITATION = HEP-PH/0610242;%%
\bibitem [{\citenamefont {Catani}\ \emph {et~al.}(1991)\citenamefont {Catani},
  \citenamefont {Dokshitzer}, \citenamefont {Olsson}, \citenamefont {Turnock},\
  and\ \citenamefont {Webber}}]{Catani:1991hj}%
  \BibitemOpen
  \bibfield  {author} {\bibinfo {author} {\bibfnamefont {S.}~\bibnamefont
  {Catani}}, \bibinfo {author} {\bibfnamefont {Y.~L.}\ \bibnamefont
  {Dokshitzer}}, \bibinfo {author} {\bibfnamefont {M.}~\bibnamefont {Olsson}},
  \bibinfo {author} {\bibfnamefont {G.}~\bibnamefont {Turnock}}, \ and\
  \bibinfo {author} {\bibfnamefont {B.~R.}\ \bibnamefont {Webber}},\ }\href
  {\doibase 10.1016/0370-2693(91)90196-W} {\bibfield  {journal} {\bibinfo
  {journal} {Phys. Lett.}\ }\textbf {\bibinfo {volume} {B269}},\ \bibinfo
  {pages} {432} (\bibinfo {year} {1991})}\BibitemShut {NoStop}%
%%CITATION = PHLTA,B269,432;%%
\bibitem [{\citenamefont {Larkoski}\ and\ \citenamefont
  {Thaler}(2014)}]{Larkoski:2014bia}%
  \BibitemOpen
  \bibfield  {author} {\bibinfo {author} {\bibfnamefont {A.~J.}\ \bibnamefont
  {Larkoski}}\ and\ \bibinfo {author} {\bibfnamefont {J.}~\bibnamefont
  {Thaler}},\ }\href {\doibase 10.1103/PhysRevD.90.034010} {\bibfield
  {journal} {\bibinfo  {journal} {Phys.Rev.}\ }\textbf {\bibinfo {volume}
  {D90}},\ \bibinfo {pages} {034010} (\bibinfo {year} {2014})},\ \Eprint
  {http://arxiv.org/abs/1406.7011} {arXiv:1406.7011 [hep-ph]} \BibitemShut
  {NoStop}%
%%CITATION = ARXIV:1406.7011;%%
\bibitem [{\citenamefont {Bauer}\ \emph {et~al.}(2001)\citenamefont {Bauer},
  \citenamefont {Fleming}, \citenamefont {Pirjol},\ and\ \citenamefont
  {Stewart}}]{Bauer:2000yr}%
  \BibitemOpen
  \bibfield  {author} {\bibinfo {author} {\bibfnamefont {C.~W.}\ \bibnamefont
  {Bauer}}, \bibinfo {author} {\bibfnamefont {S.}~\bibnamefont {Fleming}},
  \bibinfo {author} {\bibfnamefont {D.}~\bibnamefont {Pirjol}}, \ and\ \bibinfo
  {author} {\bibfnamefont {I.~W.}\ \bibnamefont {Stewart}},\ }\href {\doibase
  10.1103/PhysRevD.63.114020} {\bibfield  {journal} {\bibinfo  {journal}
  {Phys.Rev.}\ }\textbf {\bibinfo {volume} {D63}},\ \bibinfo {pages} {114020}
  (\bibinfo {year} {2001})},\ \Eprint {http://arxiv.org/abs/hep-ph/0011336}
  {arXiv:hep-ph/0011336 [hep-ph]} \BibitemShut {NoStop}%
%%CITATION = HEP-PH/0011336;%%
\bibitem [{\citenamefont {Bauer}\ and\ \citenamefont
  {Stewart}(2001)}]{Bauer:2001ct}%
  \BibitemOpen
  \bibfield  {author} {\bibinfo {author} {\bibfnamefont {C.~W.}\ \bibnamefont
  {Bauer}}\ and\ \bibinfo {author} {\bibfnamefont {I.~W.}\ \bibnamefont
  {Stewart}},\ }\href {\doibase 10.1016/S0370-2693(01)00902-9} {\bibfield
  {journal} {\bibinfo  {journal} {Phys.Lett.}\ }\textbf {\bibinfo {volume}
  {B516}},\ \bibinfo {pages} {134} (\bibinfo {year} {2001})},\ \Eprint
  {http://arxiv.org/abs/hep-ph/0107001} {arXiv:hep-ph/0107001 [hep-ph]}
  \BibitemShut {NoStop}%
%%CITATION = HEP-PH/0107001;%%
\bibitem [{\citenamefont {Bauer}\ \emph
  {et~al.}(2002{\natexlab{a}})\citenamefont {Bauer}, \citenamefont {Pirjol},\
  and\ \citenamefont {Stewart}}]{Bauer:2001yt}%
  \BibitemOpen
  \bibfield  {author} {\bibinfo {author} {\bibfnamefont {C.~W.}\ \bibnamefont
  {Bauer}}, \bibinfo {author} {\bibfnamefont {D.}~\bibnamefont {Pirjol}}, \
  and\ \bibinfo {author} {\bibfnamefont {I.~W.}\ \bibnamefont {Stewart}},\
  }\href {\doibase 10.1103/PhysRevD.65.054022} {\bibfield  {journal} {\bibinfo
  {journal} {Phys.Rev.}\ }\textbf {\bibinfo {volume} {D65}},\ \bibinfo {pages}
  {054022} (\bibinfo {year} {2002}{\natexlab{a}})},\ \Eprint
  {http://arxiv.org/abs/hep-ph/0109045} {arXiv:hep-ph/0109045 [hep-ph]}
  \BibitemShut {NoStop}%
%%CITATION = HEP-PH/0109045;%%
\bibitem [{\citenamefont {Bauer}\ \emph
  {et~al.}(2002{\natexlab{b}})\citenamefont {Bauer}, \citenamefont {Fleming},
  \citenamefont {Pirjol}, \citenamefont {Rothstein},\ and\ \citenamefont
  {Stewart}}]{Bauer:2002nz}%
  \BibitemOpen
  \bibfield  {author} {\bibinfo {author} {\bibfnamefont {C.~W.}\ \bibnamefont
  {Bauer}}, \bibinfo {author} {\bibfnamefont {S.}~\bibnamefont {Fleming}},
  \bibinfo {author} {\bibfnamefont {D.}~\bibnamefont {Pirjol}}, \bibinfo
  {author} {\bibfnamefont {I.~Z.}\ \bibnamefont {Rothstein}}, \ and\ \bibinfo
  {author} {\bibfnamefont {I.~W.}\ \bibnamefont {Stewart}},\ }\href {\doibase
  10.1103/PhysRevD.66.014017} {\bibfield  {journal} {\bibinfo  {journal}
  {Phys.Rev.}\ }\textbf {\bibinfo {volume} {D66}},\ \bibinfo {pages} {014017}
  (\bibinfo {year} {2002}{\natexlab{b}})},\ \Eprint
  {http://arxiv.org/abs/hep-ph/0202088} {arXiv:hep-ph/0202088 [hep-ph]}
  \BibitemShut {NoStop}%
%%CITATION = HEP-PH/0202088;%%
\bibitem [{\citenamefont {Bertolini}\ \emph {et~al.}(2014)\citenamefont
  {Bertolini}, \citenamefont {Chan},\ and\ \citenamefont
  {Thaler}}]{Bertolini:2013iqa}%
  \BibitemOpen
  \bibfield  {author} {\bibinfo {author} {\bibfnamefont {D.}~\bibnamefont
  {Bertolini}}, \bibinfo {author} {\bibfnamefont {T.}~\bibnamefont {Chan}}, \
  and\ \bibinfo {author} {\bibfnamefont {J.}~\bibnamefont {Thaler}},\ }\href
  {\doibase 10.1007/JHEP04(2014)013} {\bibfield  {journal} {\bibinfo  {journal}
  {JHEP}\ }\textbf {\bibinfo {volume} {1404}},\ \bibinfo {pages} {013}
  (\bibinfo {year} {2014})},\ \Eprint {http://arxiv.org/abs/1310.7584}
  {arXiv:1310.7584 [hep-ph]} \BibitemShut {NoStop}%
%%CITATION = ARXIV:1310.7584;%%
\bibitem [{\citenamefont {Bertolini}\ \emph {et~al.}(2015)\citenamefont
  {Bertolini}, \citenamefont {Thaler},\ and\ \citenamefont
  {Walsh}}]{Bertolini:2015pka}%
  \BibitemOpen
  \bibfield  {author} {\bibinfo {author} {\bibfnamefont {D.}~\bibnamefont
  {Bertolini}}, \bibinfo {author} {\bibfnamefont {J.}~\bibnamefont {Thaler}}, \
  and\ \bibinfo {author} {\bibfnamefont {J.~R.}\ \bibnamefont {Walsh}},\ }\href
  {\doibase 10.1007/JHEP05(2015)008} {\bibfield  {journal} {\bibinfo  {journal}
  {JHEP}\ }\textbf {\bibinfo {volume} {05}},\ \bibinfo {pages} {008} (\bibinfo
  {year} {2015})},\ \Eprint {http://arxiv.org/abs/1501.01965} {arXiv:1501.01965
  [hep-ph]} \BibitemShut {NoStop}%
%%CITATION = ARXIV:1501.01965;%%
\bibitem [{\citenamefont {Korchemsky}\ and\ \citenamefont
  {Sterman}(1999)}]{Korchemsky:1999kt}%
  \BibitemOpen
  \bibfield  {author} {\bibinfo {author} {\bibfnamefont {G.~P.}\ \bibnamefont
  {Korchemsky}}\ and\ \bibinfo {author} {\bibfnamefont {G.~F.}\ \bibnamefont
  {Sterman}},\ }\href {\doibase 10.1016/S0550-3213(99)00308-9} {\bibfield
  {journal} {\bibinfo  {journal} {Nucl. Phys.}\ }\textbf {\bibinfo {volume}
  {B555}},\ \bibinfo {pages} {335} (\bibinfo {year} {1999})},\ \Eprint
  {http://arxiv.org/abs/hep-ph/9902341} {arXiv:hep-ph/9902341 [hep-ph]}
  \BibitemShut {NoStop}%
%%CITATION = HEP-PH/9902341;%%
\bibitem [{\citenamefont {Korchemsky}\ and\ \citenamefont
  {Tafat}(2000)}]{Korchemsky:2000kp}%
  \BibitemOpen
  \bibfield  {author} {\bibinfo {author} {\bibfnamefont {G.~P.}\ \bibnamefont
  {Korchemsky}}\ and\ \bibinfo {author} {\bibfnamefont {S.}~\bibnamefont
  {Tafat}},\ }\href {\doibase 10.1088/1126-6708/2000/10/010} {\bibfield
  {journal} {\bibinfo  {journal} {JHEP}\ }\textbf {\bibinfo {volume} {10}},\
  \bibinfo {pages} {010} (\bibinfo {year} {2000})},\ \Eprint
  {http://arxiv.org/abs/hep-ph/0007005} {arXiv:hep-ph/0007005 [hep-ph]}
  \BibitemShut {NoStop}%
%%CITATION = HEP-PH/0007005;%%
\bibitem [{\citenamefont {Ligeti}\ \emph {et~al.}(2008)\citenamefont {Ligeti},
  \citenamefont {Stewart},\ and\ \citenamefont {Tackmann}}]{Ligeti:2008ac}%
  \BibitemOpen
  \bibfield  {author} {\bibinfo {author} {\bibfnamefont {Z.}~\bibnamefont
  {Ligeti}}, \bibinfo {author} {\bibfnamefont {I.~W.}\ \bibnamefont {Stewart}},
  \ and\ \bibinfo {author} {\bibfnamefont {F.~J.}\ \bibnamefont {Tackmann}},\
  }\href {\doibase 10.1103/PhysRevD.78.114014} {\bibfield  {journal} {\bibinfo
  {journal} {Phys. Rev.}\ }\textbf {\bibinfo {volume} {D78}},\ \bibinfo {pages}
  {114014} (\bibinfo {year} {2008})},\ \Eprint {http://arxiv.org/abs/0807.1926}
  {arXiv:0807.1926 [hep-ph]} \BibitemShut {NoStop}%
%%CITATION = ARXIV:0807.1926;%%
\bibitem [{\citenamefont {Dokshitzer}\ and\ \citenamefont
  {Webber}(1995)}]{Dokshitzer:1995zt}%
  \BibitemOpen
  \bibfield  {author} {\bibinfo {author} {\bibfnamefont {Y.~L.}\ \bibnamefont
  {Dokshitzer}}\ and\ \bibinfo {author} {\bibfnamefont {B.~R.}\ \bibnamefont
  {Webber}},\ }\href {\doibase 10.1016/0370-2693(95)00548-Y} {\bibfield
  {journal} {\bibinfo  {journal} {Phys. Lett.}\ }\textbf {\bibinfo {volume}
  {B352}},\ \bibinfo {pages} {451} (\bibinfo {year} {1995})},\ \Eprint
  {http://arxiv.org/abs/hep-ph/9504219} {arXiv:hep-ph/9504219 [hep-ph]}
  \BibitemShut {NoStop}%
%%CITATION = HEP-PH/9504219;%%
\bibitem [{\citenamefont {Lee}\ and\ \citenamefont
  {Sterman}(2007)}]{Lee:2006nr}%
  \BibitemOpen
  \bibfield  {author} {\bibinfo {author} {\bibfnamefont {C.}~\bibnamefont
  {Lee}}\ and\ \bibinfo {author} {\bibfnamefont {G.~F.}\ \bibnamefont
  {Sterman}},\ }\href {\doibase 10.1103/PhysRevD.75.014022} {\bibfield
  {journal} {\bibinfo  {journal} {Phys. Rev.}\ }\textbf {\bibinfo {volume}
  {D75}},\ \bibinfo {pages} {014022} (\bibinfo {year} {2007})},\ \Eprint
  {http://arxiv.org/abs/hep-ph/0611061} {arXiv:hep-ph/0611061 [hep-ph]}
  \BibitemShut {NoStop}%
%%CITATION = HEP-PH/0611061;%%
\bibitem [{\citenamefont {Sjostrand}\ \emph {et~al.}(2008)\citenamefont
  {Sjostrand}, \citenamefont {Mrenna},\ and\ \citenamefont
  {Skands}}]{Sjostrand:2007gs}%
  \BibitemOpen
  \bibfield  {author} {\bibinfo {author} {\bibfnamefont {T.}~\bibnamefont
  {Sjostrand}}, \bibinfo {author} {\bibfnamefont {S.}~\bibnamefont {Mrenna}}, \
  and\ \bibinfo {author} {\bibfnamefont {P.~Z.}\ \bibnamefont {Skands}},\
  }\href {\doibase 10.1016/j.cpc.2008.01.036} {\bibfield  {journal} {\bibinfo
  {journal} {Comput.Phys.Commun.}\ }\textbf {\bibinfo {volume} {178}},\
  \bibinfo {pages} {852} (\bibinfo {year} {2008})},\ \Eprint
  {http://arxiv.org/abs/0710.3820} {arXiv:0710.3820 [hep-ph]} \BibitemShut
  {NoStop}%
%%CITATION = ARXIV:0710.3820;%%
\bibitem [{Sjo(2015)}]{Sjostrand:2014zea}%
  \BibitemOpen
  \href {\doibase 10.1016/j.cpc.2015.01.024} {\bibfield  {journal} {\bibinfo
  {journal} {Comput. Phys. Commun.}\ }\textbf {\bibinfo {volume} {191}},\
  \bibinfo {pages} {159} (\bibinfo {year} {2015})},\ \Eprint
  {http://arxiv.org/abs/1410.3012} {arXiv:1410.3012 [hep-ph]} \BibitemShut
  {NoStop}%
%%CITATION = ARXIV:1410.3012;%%
\bibitem [{\citenamefont {Giele}\ \emph {et~al.}(2008)\citenamefont {Giele},
  \citenamefont {Kosower},\ and\ \citenamefont {Skands}}]{Giele:2007di}%
  \BibitemOpen
  \bibfield  {author} {\bibinfo {author} {\bibfnamefont {W.~T.}\ \bibnamefont
  {Giele}}, \bibinfo {author} {\bibfnamefont {D.~A.}\ \bibnamefont {Kosower}},
  \ and\ \bibinfo {author} {\bibfnamefont {P.~Z.}\ \bibnamefont {Skands}},\
  }\href {\doibase 10.1103/PhysRevD.78.014026} {\bibfield  {journal} {\bibinfo
  {journal} {Phys. Rev.}\ }\textbf {\bibinfo {volume} {D78}},\ \bibinfo {pages}
  {014026} (\bibinfo {year} {2008})},\ \Eprint {http://arxiv.org/abs/0707.3652}
  {arXiv:0707.3652 [hep-ph]} \BibitemShut {NoStop}%
%%CITATION = ARXIV:0707.3652;%%
\bibitem [{\citenamefont {Giele}\ \emph {et~al.}(2011)\citenamefont {Giele},
  \citenamefont {Kosower},\ and\ \citenamefont {Skands}}]{Giele:2011cb}%
  \BibitemOpen
  \bibfield  {author} {\bibinfo {author} {\bibfnamefont {W.~T.}\ \bibnamefont
  {Giele}}, \bibinfo {author} {\bibfnamefont {D.~A.}\ \bibnamefont {Kosower}},
  \ and\ \bibinfo {author} {\bibfnamefont {P.~Z.}\ \bibnamefont {Skands}},\
  }\href {\doibase 10.1103/PhysRevD.84.054003} {\bibfield  {journal} {\bibinfo
  {journal} {Phys. Rev.}\ }\textbf {\bibinfo {volume} {D84}},\ \bibinfo {pages}
  {054003} (\bibinfo {year} {2011})},\ \Eprint {http://arxiv.org/abs/1102.2126}
  {arXiv:1102.2126 [hep-ph]} \BibitemShut {NoStop}%
%%CITATION = ARXIV:1102.2126;%%
\bibitem [{\citenamefont {Gehrmann-De~Ridder}\ \emph
  {et~al.}(2012)\citenamefont {Gehrmann-De~Ridder}, \citenamefont {Ritzmann},\
  and\ \citenamefont {Skands}}]{GehrmannDeRidder:2011dm}%
  \BibitemOpen
  \bibfield  {author} {\bibinfo {author} {\bibfnamefont {A.}~\bibnamefont
  {Gehrmann-De~Ridder}}, \bibinfo {author} {\bibfnamefont {M.}~\bibnamefont
  {Ritzmann}}, \ and\ \bibinfo {author} {\bibfnamefont {P.~Z.}\ \bibnamefont
  {Skands}},\ }\href {\doibase 10.1103/PhysRevD.85.014013} {\bibfield
  {journal} {\bibinfo  {journal} {Phys. Rev.}\ }\textbf {\bibinfo {volume}
  {D85}},\ \bibinfo {pages} {014013} (\bibinfo {year} {2012})},\ \Eprint
  {http://arxiv.org/abs/1108.6172} {arXiv:1108.6172 [hep-ph]} \BibitemShut
  {NoStop}%
%%CITATION = ARXIV:1108.6172;%%
\bibitem [{\citenamefont {Ritzmann}\ \emph {et~al.}(2013)\citenamefont
  {Ritzmann}, \citenamefont {Kosower},\ and\ \citenamefont
  {Skands}}]{Ritzmann:2012ca}%
  \BibitemOpen
  \bibfield  {author} {\bibinfo {author} {\bibfnamefont {M.}~\bibnamefont
  {Ritzmann}}, \bibinfo {author} {\bibfnamefont {D.~A.}\ \bibnamefont
  {Kosower}}, \ and\ \bibinfo {author} {\bibfnamefont {P.}~\bibnamefont
  {Skands}},\ }\href {\doibase 10.1016/j.physletb.2012.12.003} {\bibfield
  {journal} {\bibinfo  {journal} {Phys. Lett.}\ }\textbf {\bibinfo {volume}
  {B718}},\ \bibinfo {pages} {1345} (\bibinfo {year} {2013})},\ \Eprint
  {http://arxiv.org/abs/1210.6345} {arXiv:1210.6345 [hep-ph]} \BibitemShut
  {NoStop}%
%%CITATION = ARXIV:1210.6345;%%
\bibitem [{\citenamefont {Hartgring}\ \emph {et~al.}(2013)\citenamefont
  {Hartgring}, \citenamefont {Laenen},\ and\ \citenamefont
  {Skands}}]{Hartgring:2013jma}%
  \BibitemOpen
  \bibfield  {author} {\bibinfo {author} {\bibfnamefont {L.}~\bibnamefont
  {Hartgring}}, \bibinfo {author} {\bibfnamefont {E.}~\bibnamefont {Laenen}}, \
  and\ \bibinfo {author} {\bibfnamefont {P.}~\bibnamefont {Skands}},\ }\href
  {\doibase 10.1007/JHEP10(2013)127} {\bibfield  {journal} {\bibinfo  {journal}
  {JHEP}\ }\textbf {\bibinfo {volume} {10}},\ \bibinfo {pages} {127} (\bibinfo
  {year} {2013})},\ \Eprint {http://arxiv.org/abs/1303.4974} {arXiv:1303.4974
  [hep-ph]} \BibitemShut {NoStop}%
%%CITATION = ARXIV:1303.4974;%%
\bibitem [{\citenamefont {Larkoski}\ \emph
  {et~al.}(2013{\natexlab{b}})\citenamefont {Larkoski}, \citenamefont
  {Lopez-Villarejo},\ and\ \citenamefont {Skands}}]{Larkoski:2013yi}%
  \BibitemOpen
  \bibfield  {author} {\bibinfo {author} {\bibfnamefont {A.~J.}\ \bibnamefont
  {Larkoski}}, \bibinfo {author} {\bibfnamefont {J.~J.}\ \bibnamefont
  {Lopez-Villarejo}}, \ and\ \bibinfo {author} {\bibfnamefont {P.}~\bibnamefont
  {Skands}},\ }\href {\doibase 10.1103/PhysRevD.87.054033} {\bibfield
  {journal} {\bibinfo  {journal} {Phys. Rev.}\ }\textbf {\bibinfo {volume}
  {D87}},\ \bibinfo {pages} {054033} (\bibinfo {year} {2013}{\natexlab{b}})},\
  \Eprint {http://arxiv.org/abs/1301.0933} {arXiv:1301.0933 [hep-ph]}
  \BibitemShut {NoStop}%
%%CITATION = ARXIV:1301.0933;%%
\bibitem [{\citenamefont {Cacciari}\ \emph {et~al.}(2012)\citenamefont
  {Cacciari}, \citenamefont {Salam},\ and\ \citenamefont
  {Soyez}}]{Cacciari:2011ma}%
  \BibitemOpen
  \bibfield  {author} {\bibinfo {author} {\bibfnamefont {M.}~\bibnamefont
  {Cacciari}}, \bibinfo {author} {\bibfnamefont {G.~P.}\ \bibnamefont {Salam}},
  \ and\ \bibinfo {author} {\bibfnamefont {G.}~\bibnamefont {Soyez}},\ }\href
  {\doibase 10.1140/epjc/s10052-012-1896-2} {\bibfield  {journal} {\bibinfo
  {journal} {Eur.Phys.J.}\ }\textbf {\bibinfo {volume} {C72}},\ \bibinfo
  {pages} {1896} (\bibinfo {year} {2012})},\ \Eprint
  {http://arxiv.org/abs/1111.6097} {arXiv:1111.6097 [hep-ph]} \BibitemShut
  {NoStop}%
%%CITATION = ARXIV:1111.6097;%%
\bibitem [{\citenamefont {Treves}\ and\ \citenamefont
  {Panzeri}(1995)}]{treves1995upward}%
  \BibitemOpen
  \bibfield  {author} {\bibinfo {author} {\bibfnamefont {A.}~\bibnamefont
  {Treves}}\ and\ \bibinfo {author} {\bibfnamefont {S.}~\bibnamefont
  {Panzeri}},\ }\href@noop {} {\bibfield  {journal} {\bibinfo  {journal}
  {Neural Computation}\ }\textbf {\bibinfo {volume} {7}},\ \bibinfo {pages}
  {399} (\bibinfo {year} {1995})}\BibitemShut {NoStop}%
\bibitem [{\citenamefont {Panzeri}\ and\ \citenamefont
  {Treves}(1996)}]{panzeri1996analytical}%
  \BibitemOpen
  \bibfield  {author} {\bibinfo {author} {\bibfnamefont {S.}~\bibnamefont
  {Panzeri}}\ and\ \bibinfo {author} {\bibfnamefont {A.}~\bibnamefont
  {Treves}},\ }\href@noop {} {\bibfield  {journal} {\bibinfo  {journal}
  {Network: Computation in Neural Systems}\ }\textbf {\bibinfo {volume} {7}},\
  \bibinfo {pages} {87} (\bibinfo {year} {1996})}\BibitemShut {NoStop}%
\bibitem [{\citenamefont {Ellis}\ \emph {et~al.}(2010)\citenamefont {Ellis},
  \citenamefont {Vermilion}, \citenamefont {Walsh}, \citenamefont {Hornig},\
  and\ \citenamefont {Lee}}]{Ellis:2010rwa}%
  \BibitemOpen
  \bibfield  {author} {\bibinfo {author} {\bibfnamefont {S.~D.}\ \bibnamefont
  {Ellis}}, \bibinfo {author} {\bibfnamefont {C.~K.}\ \bibnamefont
  {Vermilion}}, \bibinfo {author} {\bibfnamefont {J.~R.}\ \bibnamefont
  {Walsh}}, \bibinfo {author} {\bibfnamefont {A.}~\bibnamefont {Hornig}}, \
  and\ \bibinfo {author} {\bibfnamefont {C.}~\bibnamefont {Lee}},\ }\href
  {\doibase 10.1007/JHEP11(2010)101} {\bibfield  {journal} {\bibinfo  {journal}
  {JHEP}\ }\textbf {\bibinfo {volume} {1011}},\ \bibinfo {pages} {101}
  (\bibinfo {year} {2010})},\ \Eprint {http://arxiv.org/abs/1001.0014}
  {arXiv:1001.0014 [hep-ph]} \BibitemShut {NoStop}%
%%CITATION = ARXIV:1001.0014;%%
\bibitem [{\citenamefont {Berger}\ \emph {et~al.}(2003)\citenamefont {Berger},
  \citenamefont {Kucs},\ and\ \citenamefont {Sterman}}]{Berger:2003iw}%
  \BibitemOpen
  \bibfield  {author} {\bibinfo {author} {\bibfnamefont {C.~F.}\ \bibnamefont
  {Berger}}, \bibinfo {author} {\bibfnamefont {T.}~\bibnamefont {Kucs}}, \ and\
  \bibinfo {author} {\bibfnamefont {G.~F.}\ \bibnamefont {Sterman}},\ }\href
  {\doibase 10.1103/PhysRevD.68.014012} {\bibfield  {journal} {\bibinfo
  {journal} {Phys. Rev.}\ }\textbf {\bibinfo {volume} {D68}},\ \bibinfo {pages}
  {014012} (\bibinfo {year} {2003})},\ \Eprint
  {http://arxiv.org/abs/hep-ph/0303051} {arXiv:hep-ph/0303051 [hep-ph]}
  \BibitemShut {NoStop}%
%%CITATION = HEP-PH/0303051;%%
\bibitem [{\citenamefont {Almeida}\ \emph {et~al.}(2009)\citenamefont
  {Almeida}, \citenamefont {Lee}, \citenamefont {Perez}, \citenamefont
  {Sterman}, \citenamefont {Sung},\ and\ \citenamefont
  {Virzi}}]{Almeida:2008yp}%
  \BibitemOpen
  \bibfield  {author} {\bibinfo {author} {\bibfnamefont {L.~G.}\ \bibnamefont
  {Almeida}}, \bibinfo {author} {\bibfnamefont {S.~J.}\ \bibnamefont {Lee}},
  \bibinfo {author} {\bibfnamefont {G.}~\bibnamefont {Perez}}, \bibinfo
  {author} {\bibfnamefont {G.~F.}\ \bibnamefont {Sterman}}, \bibinfo {author}
  {\bibfnamefont {I.}~\bibnamefont {Sung}}, \ and\ \bibinfo {author}
  {\bibfnamefont {J.}~\bibnamefont {Virzi}},\ }\href {\doibase
  10.1103/PhysRevD.79.074017} {\bibfield  {journal} {\bibinfo  {journal} {Phys.
  Rev.}\ }\textbf {\bibinfo {volume} {D79}},\ \bibinfo {pages} {074017}
  (\bibinfo {year} {2009})},\ \Eprint {http://arxiv.org/abs/0807.0234}
  {arXiv:0807.0234 [hep-ph]} \BibitemShut {NoStop}%
%%CITATION = ARXIV:0807.0234;%%
\bibitem [{\citenamefont {{Kraskov}}\ \emph {et~al.}(2004)\citenamefont
  {{Kraskov}}, \citenamefont {{St{\"o}gbauer}},\ and\ \citenamefont
  {{Grassberger}}}]{2004PhRvE..69f6138K}%
  \BibitemOpen
  \bibfield  {author} {\bibinfo {author} {\bibfnamefont {A.}~\bibnamefont
  {{Kraskov}}}, \bibinfo {author} {\bibfnamefont {H.}~\bibnamefont
  {{St{\"o}gbauer}}}, \ and\ \bibinfo {author} {\bibfnamefont {P.}~\bibnamefont
  {{Grassberger}}},\ }\href {\doibase 10.1103/PhysRevE.69.066138} {\bibfield
  {journal} {\bibinfo  {journal} {\pre}\ }\textbf {\bibinfo {volume} {69}},\
  \bibinfo {eid} {066138} (\bibinfo {year} {2004})},\ \Eprint
  {http://arxiv.org/abs/cond-mat/0305641} {cond-mat/0305641} \BibitemShut
  {NoStop}%
\end{thebibliography}%

\end{document}